\documentclass[cleveref,a4paper,numberwithinsect,nolineno]{lipics-v2021}
\usepackage{cite} 
\usepackage{amsmath}
\usepackage{amssymb}  
\usepackage{xspace}
\usepackage{cleveref}
\usepackage{enumerate}
\usepackage{complexity}
\usepackage{tikz}
\tikzstyle{every picture} = [>=latex]
\usetikzlibrary{calc,arrows,decorations.markings,decorations.pathreplacing,quotes,fit,shapes,positioning}

 \usepackage[appendix=inline]{apxproof}
\ifx\proof\inlineproof
  \let\apxmark\relax  
\else
  \def\apxmark{{\sf\bfseries$\!\!$*~}}
\fi
\newtheoremrep{theorem2}[theorem]{Theorem}
\newtheoremrep{corollary2}[theorem]{Corollary}
\newtheoremrep{lemma2}[theorem]{Lemma}        

\theoremstyle{remark}
\newtheorem{problem}[theorem]{Problem}

\newif\ifblindre

\def\ca#1{\mathcal{#1}}

\def\crd{\operatorname{cr}} 
\def\crg{\operatorname{cr}} 
\def\cra{\operatorname{cr}\mbox{$_{\!\rm a}$}} 

\tikzstyle{vertex}=[circle, fill, inner sep=1pt]
\usetikzlibrary{patterns,patterns.meta,shapes,backgrounds,positioning}

\hideLIPIcs\nolinenumbers

\begin{document}
\title{Complexity of Anchored Crossing Number and Crossing Number of Almost Planar Graphs}
\titlerunning{Anchored and Almost Planar Crossing Number}

\ifblindre
\author{Anonymous Author(s)}{~~~}{}{}{}
\authorrunning{Anonymous Author(s)}
\else
\author{Petr Hlin\v{e}n\'y}{Masaryk University, Brno, Czech republic}{hlineny@fi.muni.cz}{https://orcid.org/0000-0003-2125-1514}{}
\authorrunning{P.~Hlin\v{e}n\'y}

\fi


\maketitle              
\begin{abstract}
We deal with the problem of computing the exact crossing number of almost planar graphs
and the closely related problem of computing the exact anchored crossing number of a pair of planar graphs.
It was shown by [Cabello and Mohar, 2013] that both problems are NP-hard; although they required an unbounded number 
of high-degree vertices (in the first problem) or an unbounded number of anchors (in the second problem) to prove their result.
Somehow surprisingly, only {\em three} vertices of degree greater than~$3$, or only three anchors, are sufficient
to maintain hardness of these problems, as we prove here.
The new result also improves the previous result on hardness of joint crossing number on surfaces by [Hlin\v{e}n\'y and Salazar, 2015].
Our result is best possible in the anchored case since the anchored crossing number of a pair of planar graphs 
with two anchors each is trivial, and close to being best possible in the almost planar case since the crossing number
is polytime computable for almost planar graphs of maximum degree $3$ [Riskin 1996, Cabello and Mohar~2011].
The complexity of crossing number of almost planar graphs with one or two vertices of degree greater than $3$ is, interestingly, still wide open.
\end{abstract}

\section{Introduction}\label{sec:intro}

Determining the {\em crossing number}, i.e.\ the smallest possible number of pairwise transverse intersections 
(called {\em crossings}) of edges in a drawing in the plane, of a graph is among the most important optimization problems in topological graph theory.
As such its general computational complexity is well-researched.
Probably most famously, it is known that graphs with crossing number~$0$, i.e.\ {\em planar graphs}, can be recognized in linear time~\cite{HopcroftT74,WeiKuanW99}.
On the other hand, computing the crossing number of a graph in general is \NP-hard, even in very restricted settings~\cite{GareyJ83,Hlineny06,DBLP:journals/algorithmica/PelsmajerSS11}, and also \APX-hard~\cite{Cabello13}.
It is rather surprising that that problem stays hard even for {\em almost planar graphs}, which are the graphs composed of a planar graph and one more edge~\cite{DBLP:journals/siamcomp/CabelloM13}.

In this paper we are particularly attracted by the last mentioned problem of computing the exact crossing number of almost planar graphs
(alternatively called near-planar graphs, e.g.~in~\cite{DBLP:journals/siamcomp/CabelloM13}).
Closely related problems were in the focus of numerous papers including 
\cite{zbMATH00881174,DBLP:journals/algorithmica/GutwengerMW05,DBLP:conf/gd/HlinenyS06,DBLP:journals/informaticaSI/Mohar06,DBLP:journals/algorithmica/CabelloM11,DBLP:journals/siamcomp/CabelloM13,DBLP:conf/compgeom/HlinenyD16}.
We shall denote an almost planar graph by $G+e$ where $G$ is a planar graph and $e$ a (new) edge with both ends in~$V(G)$.
The problem to compute the crossing number of $G+e$ is polynomial-time solvable if $G$ is planar of maximum degree $3$, by Riskin~\cite{zbMATH00881174} and
by Cabello and Mohar~\cite{DBLP:journals/algorithmica/CabelloM11}
-- as they proved, actually, the linear-time algorithm for edge insertion by Gutwenger et al.~\cite{DBLP:journals/algorithmica/GutwengerMW05} can be used to solve it.
On the other hand, the same problem to compute the crossing number of $G+e$ with planar $G$ and without restricting the degrees is \NP-hard,
as proved in Cabello and Mohar~\cite{DBLP:journals/siamcomp/CabelloM13}.

\begin{figure}[t]
{\centering\hfill a)
\begin{tikzpicture}[scale=0.8]\small
\draw[gray, ultra thick, densely dotted] (0,0) ellipse (60pt and 60pt);
\tikzstyle{every node}=[draw, color=black, thick, shape=circle, inner sep=1.3pt, fill=black]
\node (a1) at (-1,1.85) {}; \node (a2) at (1,1.85) {};
\node (a3) at (-1,-1.85) {}; \node (a4) at (1,-1.85) {};
\tikzstyle{every node}=[draw, color=black, thick, shape=circle, inner sep=1pt, fill=black]
\node (v1) at (-0.8,0.8) {}; \node (v2) at (0.8,0.8) {};
\node (v3) at (-0.8,-0.8) {}; \node (v4) at (0.8,-0.8) {};
\node (v5) at (0,0.8) {}; \node (v6) at (0,-0.8) {};
\draw (a1)--(v1)--(v2)--(a1);
\draw (a2)--(v1)--(v2)--(a2);
\draw (a3)--(v3)--(v4)--(a3);
\draw (a4)--(v3)--(v4)--(a4);
\draw (v1)--(v3)--(v5)--(v6)--(v2)--(v4);
\end{tikzpicture}
\hfill\quad b)
\begin{tikzpicture}[scale=0.8]\small
\draw[gray, ultra thick, densely dotted] (0,0) ellipse (60pt and 60pt);
\tikzstyle{every node}=[draw, color=black, thick, shape=circle, inner sep=1.3pt, fill=black]
\node (a1) at (-2.1,0) {}; \node (a2) at (2.1,0) {};
\node (a3) at (0,-2.1) {}; \node (a4) at (0,2.1) {};
\tikzstyle{every node}=[draw, color=black, thick, shape=circle, inner sep=1pt, fill=black]
\node (v1) at (-1.5,-0.5) {}; \node (v2) at (-1.5,0.5) {};
\node (v3) at (-0.7,-0.5) {}; \node (v4) at (-0.7,0.5) {};
\node (v5) at (-1.2,0) {}; 
\draw (a1)--(v1)--(v3) (v4)--(v2)--(a1)--(v5)--(v3)--(v4)--(v5);
\node (v6) at (1.2,0) {}; 
\node (v7) at (1.2,-0.5) {}; \node (v8) at (1.2,0.5) {}; 
\draw (a2)--(v7)--(v8)--(a2)--(v6);
\draw (v3)--(v7) (v4)--(v8);
\node (w1) at (-0.5,-1.4) {}; \node (w2) at (0.5,-1.4) {};
\node (w3) at (0,-1) {}; 
\node (w4) at (-0.5,1) {}; \node (w5) at (0.5,1) {};
\node (w6) at (0,1.4) {}; 
\draw (a3)--(w1)--(w3)--(w2)--(a3)--(w3);
\draw (a4)--(w4)--(w5)--(a4)--(w6)--(w4) (w6)--(w5);
\draw (w3)--(w4) (w2)--(w5);
\end{tikzpicture}
\hfill}

\caption{a) An example of an anchored graph. Although the graph itself is planar, its anchored crossing number equals~$2$.
	b) Another example made of a disjoint union of two anchored planar graphs. Its anchored crossing number equals~$4$.
}\label{fig:anchbasic}
\end{figure}
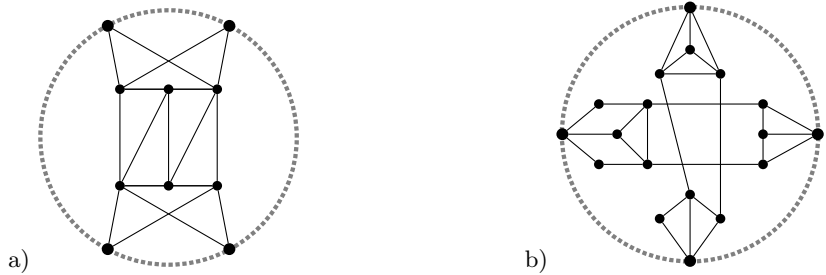

\subparagraph*{Hardness of the anchored crossing number. }
The hardness proof in~\cite{DBLP:journals/siamcomp/CabelloM13} importantly builds on the concept of {\em anchored crossing number}, which is of independent interest.
The anchored crossing number $\cra(H)$ of an anchored graph $H$ is defined the same way as the ordinary crossing number, with an additional restriction that allowed drawings
of $H$ (called {\em anchored drawings}) must be contained in a disk such that prescribed vertices of $H$ (the {\em anchors}) 
are placed (``anchored'') in prescribed distinct points on the disk boundary.
See \Cref{fig:anchbasic}\,a).
A graph $H$ is {\em anchored planar} if the anchored crossing number of $H$ is~$0$ (note that a planar graph may not be anchored planar for some/all selections of the anchor vertices).

\begin{theorem}[Cabello and Mohar~\cite{DBLP:journals/siamcomp/CabelloM13}]\label{thm:anchorNPH}
Computing the anchored crossing number of an anchored graph $H$ is \NP-hard, even if $H$ is the union of two vertex-disjoint anchored planar graphs.
\end{theorem}

The hard instances $H$ in \Cref{thm:anchorNPH} have an additional property (cf.~\cite[Corollary~2.7]{DBLP:journals/siamcomp/CabelloM13}), which will be important later:
Each of the two disjoint anchored planar subgraphs forming $H$ has a unique anchored planar drawing (essentially), and every optimal
solution to the anchored crossing number of $H$ is a union of these unique anchored planar drawings.

The hardness construction in \cite{DBLP:journals/siamcomp/CabelloM13} had used an unbounded number of anchor vertices,
but Hlin\v{e}n\'y and Salazar \cite{DBLP:conf/isaac/HlinenyS15} noted that the conclusion of \Cref{thm:anchorNPH}
remains true if the graph $H$ moreover has a bounded number (namely at most~$16$) anchor vertices.
We now discuss further details in this direction.

Let the anchored graph $H$ from \Cref{thm:anchorNPH} be written as a disjoint union $H=H_1\cup H_2$, where each $H_i$, $i=1,2$, is anchored planar with $a_i>0$ anchors.
Observe that if \mbox{$\min\{a_1,a_2\}=1$}, then, trivially, $H$ is anchored planar as well.
If $\min\{a_1,a_2\}=2=a_1$ (up to symmetry), then we can efficiently compute the smallest edge cut between the two anchors
of $H_1$, and multiply it by the smallest edge cut in $H_2$ between the corresponding two groups (as separated by~the anchors of $H_1$) of the anchors of $H_2$.
This product equals, again quite trivially, the anchored crossing number of $H$.
See \Cref{fig:anchbasic}\,b).
This leaves $a_1=a_2=3$ as the simplest possibly nontrivial case of the special anchored crossing number problem.

We show that the latter case is already hard in our main result proved later in the paper:

\begin{theorem}\label{thm:main3}
Let $H$ be an anchored graph such that $H=H_1\cup H_2$, where $H_1$ and $H_2$ are vertex-disjoint connected anchored planar graphs, each with $3$ anchors,
and given alongside with anchored planar drawings $\ca D_1$ and $\ca D_2$, respectively.
Moreover, assume that $H$ is such that in every optimal solution to the anchored crossing number of $H$, the subdrawing of $H_i$, $i\in\{1,2\}$,
is equivalent (homeomorphic) up to permutations of parallel edges to the given drawing~$\ca D_i$.
Then it is \NP-hard to compute the anchored crossing number of~$H$.
\end{theorem}
Note that parallel edges do not play any essential role in the crossing-number context, since they can be subdivided to make the graph simple,
or modelled by integer-weighted simple edges as we do here later from \Cref{sec:prelim}.

\subparagraph*{Back to the crossing number of almost planar graphs. }
The special conditions on hard instances formulated in \Cref{thm:main3} have an interesting consequence which we informally outline next.
We first recall a trick introduced in~\cite{DBLP:journals/siamcomp/CabelloM13}:
Having an instance $H=H_1\cup H_2$ of the anchored crossing number problem as in \Cref{thm:main3}, we can construct a planar graph $G_0:=H\cup C^+$
where $C^+$ is a (multi)cycle on the $6$ anchor vertices of $H$ in the natural cyclic order, with ``sufficiently many'' parallel edges between consecutive pairs of the vertices of~$C^+$.
See \Cref{fig:anchtoalmost}.
Now we choose vertices $v_i\in V(H_i)\setminus V(C^+)$, $i=1,2$, and add a new edge $f=v_1v_2$ to $G_0$.
Then any optimal solution to the (now ordinary) crossing number of $G_0+f$
must leave $C^+$ uncrossed, and so it is actually an anchored drawing of $G_0+f$ with the anchors~$V(C^+)$.

Assuming we choose in the previous $v_1$ and $v_2$ such that the anchored crossing number of $G_0+v_1v_2$ equals that of~$H$
(which is quite easy under the assumption of given drawings $\ca D_1$ and $\ca D_2$ in \Cref{thm:main3}), we continue as in \Cref{fig:anchtoalmost}\,c).
We modify the graph $G_0$ into $\bar G_0$ by blowing up every vertex of $V(H_1)$ and every vertex of $V(H_2)\setminus V(C^+)$ into a ``sufficiently large'' cubic grid.
Then every vertex~of~$\bar G_0$, except the three anchor vertices of $H_2$, is of degree at most~$3$, and $\bar G_0$ is still planar 
(since flexibility of the anchors of $H_2$ allows to flip modified $H_2$ ``inside~out'' within $\bar G_0$).
Furthermore, the crossing number of $\bar G_0+v_1v_2$ equals the anchored crossing number of~$H$.
Formal details to be provided in the later proof. 

\pgfdeclarepattern{name=Bricks,
  parameters={\brickwidth,\brickheight,\brickangle,\bricklinewidth},
  bottom left={\pgfpoint{-.1pt}{-.1pt}},
  top right={\pgfpoint{\brickwidth+.1pt}{\brickheight+.1pt}},
  tile size={\pgfpoint{\brickwidth}{\brickheight}},
  tile transformation={\pgftransformrotate{\brickangle}},
  code={
    \pgfsetlinewidth{\bricklinewidth}
    \pgfpathmoveto{\pgfpoint{-.1pt}{0pt}}
    \pgfpathlineto{\pgfpoint{\brickwidth+.1pt}{0pt}}
    \pgfpathmoveto{\pgfpoint{-.1pt}{0.5*\brickheight}}
    \pgfpathlineto{\pgfpoint{\brickwidth+.1pt}{0.5*\brickheight}}
	\pgfpathmoveto{\pgfpoint{0.25*\brickwidth}{0pt}}
    \pgfpathlineto{\pgfpoint{0.25*\brickwidth}{0.5*\brickheight}}
	\pgfpathmoveto{\pgfpoint{0.75*\brickwidth}{0.5*\brickheight}}
    \pgfpathlineto{\pgfpoint{0.75*\brickwidth}{\brickheight}}
    \pgfusepath{stroke}
} }
\tikzset{/pgf/pattern keys/.cd,
brick width/.store in=\brickwidth,
brick height/.store in=\brickheight,
brick angle/.store in=\brickangle,
brick line width/.store in=\bricklinewidth, 
brick width=1em,
brick height=1em,
brick angle=0pt,
brick line width=.5pt,
}

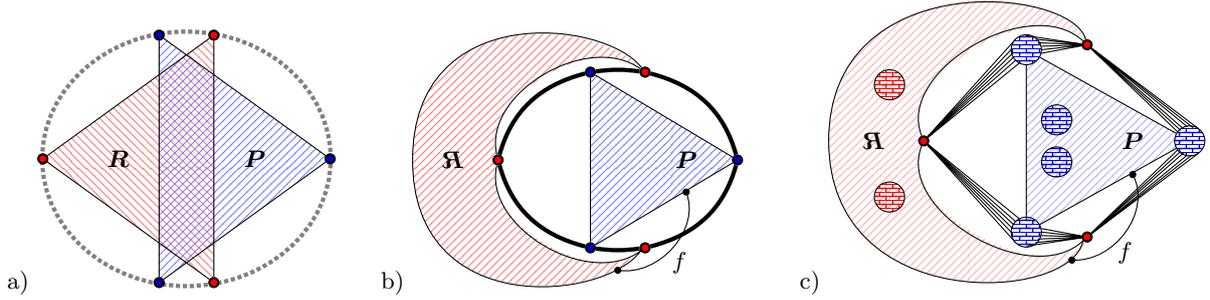
\begin{figure}[t]
\centering a)\hspace*{-1ex}
\begin{tikzpicture}[xscale=0.8,yscale=0.8]\small
\draw[gray, ultra thick, densely dotted] (0,0) ellipse (60pt and 60pt);
\tikzstyle{every node}=[draw, color=black, thick, shape=circle, inner sep=1.2pt, fill=red]
\draw[draw=none, pattern=north west lines, pattern color=red!50!white] (-2.1,0)--(0.4,-2.05)--(0.4,2.05)--(-2.1,0);
\draw[draw=none, pattern=north east lines, pattern color=blue!50!white] (2.1,0)--(-0.4,-2.05)--(-0.4,2.05)--(2.1,0);
\node (a1) at (-2.1,0) {};
\node (a2) at (0.4,-2.05) {}; \node (a3) at (0.4,2.05) {};
\tikzstyle{every node}=[draw, color=black, thick, shape=circle, inner sep=1.2pt, fill=blue!80!black]
\node (b1) at (2.1,0) {};
\node (b2) at (-0.4,-2.05) {}; \node (b3) at (-0.4,2.05) {};
\tikzstyle{every path}=[draw, color=black]
\draw (a1)--(a2)--(a3)--(a1);
\draw (b1)--(b2)--(b3)--(b1);
\node[draw=none,fill=none] at (1,0) {{\boldmath$P$}};
\node[draw=none,fill=none] at (-1,0) {{\boldmath$R$}};
\end{tikzpicture}
\hfill b)\hspace*{-1ex}
\begin{tikzpicture}[xscale=0.8,yscale=0.9]\small
\path [use as bounding box] (-3,-2) rectangle (2,2.6);
\tikzstyle{every node}=[draw, color=black, thick, shape=circle, inner sep=1.2pt, fill=red]
\draw[draw=none, pattern=north east lines, pattern color=blue!50!white] (1.75,0)--(-0.4,-1.37)--(-0.4,1.37)--(1.75,0);
\draw[draw=none, pattern=north east lines, pattern color=red!50!white] (-1.75,0) to[bend right=80] (0.4,-1.37)
	.. controls (-0.2,-2.2) and (-3,-2.5) .. (-3,0) .. controls (-3,2.5) and (-0.2,2.2) .. (0.4,1.37) to[bend right=80] (-1.75,0);
\node (a1) at (-1.75,0) {};
\node (a2) at (0.4,-1.37) {}; \node (a3) at (0.4,1.37) {};
\tikzstyle{every node}=[draw, color=black, thick, shape=circle, inner sep=1.2pt, fill=blue!80!black]
\node (b1) at (1.75,0) {};
\node (b2) at (-0.4,-1.37) {}; \node (b3) at (-0.4,1.37) {};
\tikzstyle{every path}=[draw, color=black]
\draw (b1)--(b2)--(b3)--(b1);
\draw (a1) to[bend right=70] (a2) .. controls (-0.2,-2.2) and (-3,-2.5) .. (-3,0) .. controls (-3,2.5) and (-0.2,2.2) .. (a3) to[bend right=70] (a1);
\tikzstyle{every node}=[draw, color=black, shape=circle, inner sep=0.8pt, fill=black]
\draw (0,-1.72) node {} to[bend right=60] (1,-0.5) node {};
\tikzstyle{every path}=[draw, color=black, ultra thick]
\draw (a1) to[bend right] (b2) (b2) to[bend right=8] (a2) (a2) to[bend right] (b1);
\draw (b1) to[bend right] (a3) (a3) to[bend right=8] (b3) (b3) to[bend right] (a1);
\node[draw=none,fill=none] at (1,0) {{\boldmath$P$}};
\node[draw=none,fill=none] at (0.9,-1.6) {$f$};
\node[draw=none,fill=none, xscale=-1] at (-2.4,0) {{\boldmath$R$}};
\end{tikzpicture}
\hfill c)\hspace*{-1ex}
\begin{tikzpicture}[xscale=0.9,yscale=0.9]\small
\path [use as bounding box] (-3,-2.1) rectangle (2,2.4);
\tikzstyle{every node}=[draw, color=black, thick, shape=circle, inner sep=1.2pt, fill=red]
\draw[draw=none, pattern=north east lines, pattern color=blue!30!white] (1.75,0)--(-0.4,-1.3)--(-0.4,1.3)--(1.75,0);
\draw[draw=none, pattern=north east lines, pattern color=red!30!white] (-1.75,0) to[bend right=90] (0.4,-1.37)
	.. controls (0.2,-2.3) and (-3,-2.5) .. (-3,0) .. controls (-3,2.5) and (0.2,2.3) .. (0.4,1.37) to[bend right=90] (-1.75,0);
\node (a1) at (-1.75,0) {};
\node (a2) at (0.4,-1.37) {}; \node (a3) at (0.4,1.37) {};
\tikzstyle{every node}=[draw, color=black, shape=circle, inner sep=4pt, pattern color=blue,
		pattern={Bricks[brick width=5pt,brick height=3pt]}]
\node (xx) at (0,-0.3) {}; \node (xx) at (0,0.3) {};
\node (b1) at (1.75,0) {};
\node (b2) at (-0.4,-1.3) {}; \node (b3) at (-0.4,1.3) {};
\node[pattern color=red] (xx) at (-2.2,-0.8) {};
\node[pattern color=red] (xx) at (-2.2,0.8) {};
\tikzstyle{every path}=[draw, color=black]
\draw (b1)--(b2)--(b3)--(b1);
\draw (a1) to[bend right=78] (a2) .. controls (0.2,-2.3) and (-3,-2.5) .. (-3,0) .. controls (-3,2.5) and (0.2,2.2) .. (a3) to[bend right=78] (a1);
\tikzstyle{every node}=[draw, color=black, shape=circle, inner sep=0.8pt, fill=black]
\draw (0.2,-1.7) node {} to[bend right=60] (1,-0.48) node {};
\tikzstyle{every path}=[draw, color=black]
\draw (a1)--(-0.54,-1.45) (a1)--(-0.6,-1.3) (a1)--(-0.58,-1.25) (a1)--(-0.56,-1.2) (a1)--(-0.54,-1.15) (a1)--(-0.48,-1.12);
\draw (a1)--(-0.54,1.45) (a1)--(-0.6,1.3) (a1)--(-0.58,1.25) (a1)--(-0.56,1.2) (a1)--(-0.54,1.15) (a1)--(-0.48,1.12);
\draw (a2)--(-0.4,-1.5) (a2)--(-0.28,-1.45) (a2)--(-0.25,-1.4) (a2)--(-0.2,-1.35) (a2)--(-0.2,-1.3) (a2)--(-0.2,-1.25);
\draw (a2)--(1.86,-0.18) (a2)--(1.78,-0.2) (a2)--(1.72,-0.2) (a2)--(1.67,-0.17) (a2)--(1.63,-0.15) (a2)--(1.59,-0.13);
\draw (a3)--(-0.4,1.5) (a3)--(-0.28,1.45) (a3)--(-0.25,1.4) (a3)--(-0.2,1.35) (a3)--(-0.2,1.3) (a3)--(-0.2,1.25);
\draw (a3)--(1.86,0.18) (a3)--(1.78,0.2) (a3)--(1.72,0.2) (a3)--(1.67,0.17) (a3)--(1.63,0.15) (a3)--(1.59,0.13);
\node[draw=none,fill=none] at (1,0) {{\boldmath$P$}};
\node[draw=none,fill=none] at (0.9,-1.6) {$f$};
\node[draw=none,fill=none, xscale=-1] at (-2.4,0) {{\boldmath$R$}};
\end{tikzpicture}
\smallskip

\caption{An illustration to Corollary~\ref{cor:main3}; how to translate the problem of the anchored crossing number (of a suitable instance as in \Cref{thm:main3})
	to that of the ordinary crossing number of an almost planar graph.
	a)~The depicted instance consists of a disjoint union of two (blue $P$ and red $R$) anchored planar graphs.
	b)~Turning the previous into an almost planar instance of the ordinary crossing number problem, with a ``heavy'' cycle
	on the former anchor vertices, a ``flipped~out'' subdrawing of the component $R$,
	and an added edge $f$ which effectively forces $P$ and $R$ to stay together inside the~cycle in an optimal drawing.
	c)~Blowing up every vertex except the 3 former red anchors into a large cubic grid (a~wall), which keeps the drawing properties
	and~stays~almost~planar.
}\label{fig:anchtoalmost}
\end{figure}

Hence, in contrast to the efficiently computable case of the crossing number $\crg(G+e)$ where $G$ is planar of maximum degree~$3$ 
\cite{DBLP:journals/algorithmica/CabelloM11}, we prove the following: 
\begin{corollary2rep}[of \Cref{thm:main3}]\label{cor:main3}\apxmark
Let $G$ be a planar graph such that at most three vertices of $G$ are of degree greater than~$3$, and $u,v\in V(G)$.
Then it is \NP-hard to compute the crossing number of the almost planar graph~$G+uv$.
\end{corollary2rep}

\begin{proof}
As sketched in the main body of the paper and illustrated in \Cref{fig:anchtoalmost};
we can take an instance $H=H_1\cup H_2$ of anchored crossing number as in \Cref{thm:main3},
and construct a planar graph $G_0:=H\cup C^+$ where $C^+$ is a (multi)cycle on the $6$ anchor vertices of $H$ 
in the natural cyclic order, with $m$ parallel edges between consecutive pairs of the vertices of~$C^+$.
The parameter $m$ is chosen ``sufficiently large'', e.g., $m\geq2|E(H_1)|\cdot|E(H_2)|\geq2\cra(H)+1$ in the worst-case scenario of~$H$.

Let~$e_1\in E(H_1)$ and $e_2\in E(H_2)$ be edges incident to anchor vertices and sharing a face in some
optimal solution to the anchor crossing number $\cra(H)$, and subdivide $e_i$ with a new vertex $v_i$ for $i=1,2$.
For simplicity, we use the same names $H=H_1\cup H_2$ also for the subdivided graphs.
Then $\cra(H)=\cra(H+f)$ where~$f=v_1v_2$.
We first claim that $\crg(G_0+f)=\cra(H+f)=\cra(H)$. Indeed, $\crg(G_0+f)\leq\cra(H)$ is trivial.
Assume that there is a drawing $D$ of $G_0+f$ with at most $\cra(H)<m/2$ crossings.
Then there is an uncrossed cycle $C_0\subseteq C^+$ in~$D$, and so the drawing of $C_0$ bounds a disk such that $D$ is
an anchored drawing of $H+f$ with $\crd(D)\geq\cra(H)$ crossings, proving the claim.

Second, we note that there exists a fixed rotation scheme of the edges of $G_0$ which is defined by the given
drawings $\ca D_1$ and $\ca D_2$ of \Cref{thm:main3}, and all optimal solutions to $\cra(H+f)$ by \Cref{thm:main3}, 
and hence also all optimal solutions to $\crg(G_0+f)$ by the previous paragraph, respect this rotation scheme.
Furthermore, $G_0$ has a planar drawing which respects the same rotation scheme up to mirroring and except at the three anchor vertices~$A_1\subseteq V(H_1)$
(informally, the subdrawing of $H_1$ is ``flipped out'' of the multicycle $C^+$ in the drawing of~$G_0$, as shown in \Cref{fig:anchtoalmost}).

\smallskip
The last step of the proof is to construct a planar graph $G$ from $G_0$ such that all vertices of $G$ except
$A_1\subseteq V(G)$ are of degree at most~$3$, and $\crg(G+f)=\crg(G_0+f)$.
For this we apply the technique of ``cubic grids'', used for instance in \cite{Hlineny06,DBLP:journals/algorithmica/PelsmajerSS11,Cabello13} previously.
Let a \emph{cylindrical cubic grid} (also called a cylindrical ``wall'') of height~$h$ and length~$\ell$ be the following graph~$H$:
Start with the union of $h$ cycles of length $\ell$ each, $C_1,\ldots,C_h$, such that $V(C_i)=(v^i_1,\ldots,v^i_\ell)$ in this cyclic order,
and add all edges $\{v^i_j,v^{i+1}_j\}$ where $1\leq i<m$, $1\leq j\leq\ell$ and $i+j$ is odd.
Then $H$ is planar and all vertices of $H$ are of degree $3$, except every second vertex of $C_1$ and of $C_h$. 
Let $C_1$ be called the \emph{outer cycle} of the grid~$H$.

Let $h=|E(G_0)|^2$. We do the following, as illustrated in \Cref{fig:anchtoalmost}. 
For every vertex $v\in V(G_0)\setminus A_1$ of degree $d>3$, we take a copy $H_v$ of the cylindrical cubic grid of height~$h$ and length~$2d$,
and attach every edge formerly starting in~$v$ to a distinct degree-$2$ vertex of the outer cycle of~$H_v$, in the apropriate cyclic order of the aforementioned rotation scheme of~$G_0$.
For the resulting graph $G$ we argue as follows.
First, by the assumption on the rotation scheme of $G_0$, we immediately get $\crg(G+f)\leq\crg(G_0+f)$.
Second, in any assumed optimal solution to $\crg(G+f)$ which has less than $\crg(G_0+f)<h/2$ crossings, and for any $v\in V(G_0)$ replaced by~$H_v$,
one of the $h$ cycles $C_i\subseteq H_v$ is uncrossed since the crossings affect at most $2\crg(G+f)<h$ of these cycles.
So, we may prolong every edge of $G_0$ attached to $H_v$ along a disjoint path in $H_v$ towards $C_i$,
then contract uncrossed $C_i$ into a vertex (former $v\in V(G_0)$) and obtain a drawing certifying $\crg(G_0+f)\leq\crg(G+f)$.
\end{proof}

\ifx\proof\inlineproof\else
We leave proofs of the ~~\apxmark-marked statements for the Appendix.
\fi

Corollary~\ref{cor:main3} brings a natural question of whether we can efficiently compute the exact crossing number $\crg(G+e)$ 
of almost planar graphs $G+e$ where $G$ has \emph{one or two vertices} of degree greater than~$3$, and we discuss on this in \Cref{sec:conclu}.

Consequences of \Cref{thm:main3} are not restricted only to the crossing number almost planar graphs, but include, for instance,
also the following problem \cite{DBLP:conf/isaac/HlinenyS15} of the \emph{joint crossing number in a surface}:
Given are two disjoint graphs $G_1$ and $G_2$, each one embedded on a fixed surface~$\ca S$, and the task is to find
a drawing (called simultaneous or joint) of $G_1\cup G_2$ in $\ca S$ which preserves each of the given embeddings of $G_1$ and $G_2$,
and the number of crossings between $E(G_1)$ and $E(G_2)$ is minimized.
While Hlin\v{e}n\'y and Salazar \cite{DBLP:conf/isaac/HlinenyS15} proved that this problem is hard for the orientable 
surface with $6$ handles, we easily improve the result to:
\begin{corollary2rep}[of \Cref{thm:main3}]\label{cor:joint3}\apxmark
It is \NP-hard to compute the joint crossing number of two (disjoint) graphs embedded in the triple-torus.
\end{corollary2rep}

\begin{proof}
We utilize the technique of \emph{face-anchors} from \cite{DBLP:conf/isaac/HlinenyS15};
for any $h\geq1$ and the surface with $h$ handles, this technique allows us to confine prescribed $h$ vertices of the graph $H_1$,
in the context of \Cref{thm:main3}, to lie in prescribed $h$ faces of the graph $H_2$ in any optimal joint drawing of~$H_1\cup H_2$.
Correctness of this reduction is proved for any fixed $h$ in \cite[Theorem~3.1]{DBLP:conf/isaac/HlinenyS15}, and we simply
apply it with $h=3$ onto \Cref{thm:main3}.
\end{proof}

Lastly, we note that for all mentioned problems which are \NP-hard, if the input includes an integer~$k$,
then, by standard means, it becomes \NP-complete to decide whether a solution with at most $k$ crossings is possible.

\section{Basic Definitions and Tools}\label{sec:prelim}

In this paper we consider multigraphs by default, i.e., our graphs are allowed to have multiple edges (while loops are irrelevant here),
with understanding that we can always subdivide parallel edges without changing the crossing number.

\subparagraph{Drawings. }
A \emph{drawing} $\mathcal{G}$ of a graph $G$ in the Euclidean plane $\mathbb{R}^2$ is a function that maps each vertex $v \in V(G)$ to a distinct point $\mathcal{G}(v) \in \mathbb R^2$ and each edge $e=uv \in E(G)$ to a simple open curve $\mathcal{G}(e) \subset \mathbb R^2$ with the ends $\mathcal{G}(u)$ and $\mathcal{G}(v)$.
We require that $\ca G(e)$ is disjoint from $\ca G(w)$ for all~$w\in V(G)\setminus\{u,v\}$.
In a slight abuse of notation we often identify a vertex $v$ with its image $\mathcal{G}(v)$ and an edge $e$ with~$\mathcal{G}(e)$.
Throughout the paper we will moreover assume that:
there are finitely many points which are in an intersection of two edges,
no more than two edges intersect in any single point other than a vertex,
and whenever two edges intersect in a point, they do so transversally (i.e., not tangentially).

The intersection (a point) of two edges is called a \emph{crossing} of these edges.
A drawing $\ca G$ is \emph{planar} (or a {\em plane graph}) if $\ca G$ has no crossings, and a graph is \emph{planar} if it has a planar drawing.
The number of crossings in a drawing $\ca G$ is denoted by $\crd(\ca G)$.
The {\em crossing number $\crg(G)$ of $G$} is defined as the minimum of $\crd(\ca G)$ over all drawings $\ca G$ of~$G$.

The following is a useful artifice in crossing numbers research. 
In a {\em weighted} graph, each edge is assigned a positive number (the {\em weight} or thickness of the edge, usually an integer).
Now the {\em weighted crossing number} is defined as the ordinary crossing number, but a crossing between edges $e_1$ and $e_2$,
say of weights $t_1$ and $t_2$, contributes the product $t_1\cdot t_2$ to the weighted crossing number.
For the purpose of computing the crossing number, an edge of integer weight $t$ can be equivalently replaced by a bunch of $t$ parallel edges of weights~$1$;
this is since we can easily redraw every edge of the bunch tightly along the ``cheapest'' edge of the bunch.
Hence, from now on, we will use weighted edges instead of parallel edges, and shortly say {\em crossing number} to the weighted crossing number.
(Note, though, that when we say a graph $G+e$ is almost planar, then we strictly mean that the added edge $e$ is of weight~$1$.)

\subparagraph{Anchored drawings. }
Assume now a closed disk $D\subseteq\mathbb{R}^2$.
An {\em anchored graph}~\cite{DBLP:journals/siamcomp/CabelloM13} is a pair $(H,A)$ where $A\subseteq V(H)$ is a cyclic permutation of some of its vertices;
the vertices in $A$ are called the {\em anchors of~$H$}.
An {\em anchored drawing} of $(H,A)$ is a drawing $\ca H$ of $H$ such that $\ca H\subseteq D$ and $\ca H$ intersects the boundary of $D$ exactly
in the vertices of $A$ in the prescribed cyclic order.
The {\em anchored crossing number of $(H,A)$}, denoted by $\cra(H,A)$, equals the minimum of $\crd(\ca H)$ over all anchored drawings $\ca H$ of~$(H,A)$.
An anchored graph $(H,A)$ is {\em anchored planar} if $\cra(H,A)=0$.

We shall study the following special case of an anchored graph~$(H,A)$, which we call an {\em anchored pair of planar graphs}, or shortly a {\em PP anchored graph}:
It is the case of $H=H_1\cup H_2$ such that $(H_1,A_1)$ and $(H_2,A_2)$ are vertex-disjoint anchored planar graphs, where $A_i$, $i=1,2$, 
is the restriction of the permutation $A$ to the anchor set of~$H_i$.

From \cite{DBLP:journals/siamcomp/CabelloM13} one can derive the following refined statement.
In regard of weighted graphs representing parallel edges (as mentioned above), we say that a graph $G$ contains a path $P\subseteq G$ 
of weight $t$ if every edge of $P$ in $G$ is of weight at least~$t$.

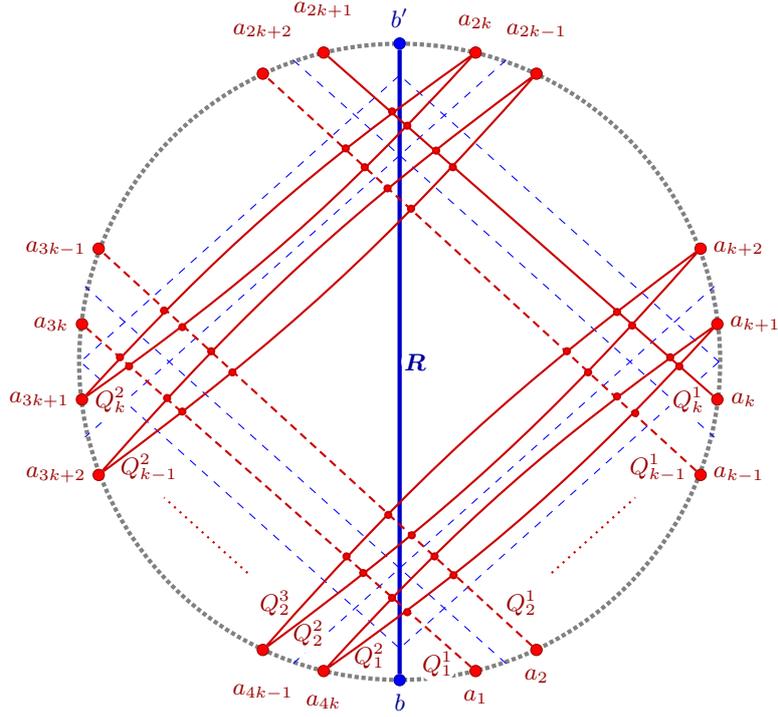
\begin{figure}[t]
\centering
\begin{tikzpicture}[scale=2.0]\small
\draw[gray, ultra thick, densely dotted] (0,0) ellipse (60pt and 60pt);
\tikzstyle{every node}=[draw, color=black!40!blue, shape=circle, inner sep=1.5pt, fill=blue]
\draw node[label=below:$b$] (b) at (0,-2.11) {}; \draw node[label=above:$b'$] (bb) at (0,2.11) {};
\draw[blue!80!black,ultra thick] (b)--(bb);
\tikzstyle{every node}=[draw, color=black!40!red, shape=circle, inner sep=1.5pt, fill=red]
\draw node[label=below:$a_1$] (a1) at (0.5,-2.05) {};
\draw node[label=below:$a_2$] (a2) at (0.9,-1.91) {};
\draw node[label=right:$a_{k-1}$] (a3) at (1.98,-0.75) {};
\draw node[label=right:$a_k$] (a4) at (2.09,-0.25) {};
\draw node[label=above:$a_{2k}$] (a8) at (0.5,2.05) {};
\draw node[label=above:$a_{2k-1}$] (a7) at (0.9,1.91) {};
\draw node[label=right:$a_{k+2}$] (a6) at (1.98,0.75) {};
\draw node[label=right:$a_{k+1}$] (a5) at (2.09,0.25) {};
\draw node[label=above:$a_{2k+1}$] (a11) at (-0.5,2.05) {};
\draw node[label=above:$a_{2k+2}$] (a12) at (-0.9,1.91) {};
\draw node[label=left:$a_{3k-1}$] (a13) at (-1.98,0.75) {};
\draw node[label=left:$a_{3k}$] (a14) at (-2.09,0.25) {};
\draw node[label=below:$a_{4k}$] (a18) at (-0.5,-2.05) {};
\draw node[label=below:$a_{4k-1}$] (a17) at (-0.9,-1.91) {};
\draw node[label=left:$a_{3k+2}$] (a16) at (-1.98,-0.75) {};
\draw node[label=left:$a_{3k+1}$] (a15) at (-2.09,-0.25) {};
\tikzstyle{every path}=[draw, color=blue, dashed]
\draw (0.7,-2)--(-2.07,0.5) (-0.7,-2)--(2.07,0.5);
\draw (-0.7,2)--(2.07,-0.5) (0.7,2)--(-2.07,-0.5);
\draw (2.1,0)--(0,-1.9)--(-2.1,0)--(0,1.9)--(2.1,0);
\tikzstyle{every node}=[draw=none, color=black!40!red, fill=white, inner sep=0pt, shape=circle]
\draw node at (0.25,-2) {$Q^1_1$} node at (0.8,-1.6) {$Q^1_2$} node at (1.7,-0.7) {$Q^1_{k-1}$} node at (1.9,-0.25) {$Q^1_{k}$};
\draw node at (-0.2,-1.95) {$Q^2_1$} node at (-0.6,-1.8) {$Q^2_2$} node at (-0.82,-1.6) {$Q^3_2$};
\draw node at (-1.65,-0.7) {$Q^2_{k-1}$} node at (-1.9,-0.25) {$Q^2_{k}$};
\tikzstyle{every node}=[draw=none, color=black!40!blue, fill=white, inner sep=0pt, shape=circle]
\draw node at (0.1,0) {{\boldmath$R$}};
\tikzstyle{every path}=[draw, color=black!20!red, thick,dotted]
\draw (1,-1.4)--(1.55,-0.9) (-1,-1.4)--(-1.55,-0.9);
\tikzstyle{every path}=[draw, color=black!20!red, thick, densely dashed]
\draw (a1)--(a14) (a2)--(a13) (a3)--(a12);
\tikzstyle{every path}=[draw, color=black!20!red, thick]
\draw (a4)--(a11);
\draw (a5) to[bend right=6] (a18) (a5) to[bend left=6] (a18);
\draw (a6) to[bend right=6] (a17) (a6) to[bend left=6] (a17);
\draw (a7) to[bend right=6] (a16) (a7) to[bend left=6] (a16);
\draw (a8) to[bend right=6] (a15) (a8) to[bend left=6] (a15);
\tikzstyle{every node}=[draw, color=black!20!red, shape=circle, inner sep=0.8pt, fill=red]
\draw node at (0.05,-1.66) {} node at (-0.05,-1.565) {} node at (0.23,-1.29) {} node at (0.355,-1.415) {};
\draw node at (0.08,-1.15) {} node at (-0.075,-1.015) {} node at (-0.35,-1.29) {} node at (-0.24,-1.4) {};
\draw node at (1.84,-0.03) {} node at (1.78,0.03) {} node at (1.53,0.24) {} node at (1.43,0.33) {};
\draw node at (1.55,-0.34) {} node at (1.43,-0.23) {} node at (1.24,-0.07) {} node at (1.1,0.07) {};
\draw node at (-0.05,1.66) {} node at (0.05,1.565) {} node at (-0.23,1.29) {} node at (-0.355,1.415) {};
\draw node at (-0.08,1.15) {} node at (0.075,1.015) {} node at (0.35,1.29) {} node at (0.24,1.4) {};
\draw node at (-1.84,0.03) {} node at (-1.78,-0.03) {} node at (-1.53,-0.24) {} node at (-1.43,-0.33) {};
\draw node at (-1.55,0.34) {} node at (-1.43,0.23) {} node at (-1.24,0.07) {} node at (-1.1,-0.07) {};
\end{tikzpicture}

\caption{A high-level illustration of the assumptions on the PP anchored graph $(H,A)$ in Theorem~\ref{thm:anchorNPHrefin};
	the drawing $\ca D_1$ of $(H_1,A_1)$ is sketched in red, and the drawing $\ca D_2$ of $(H_2,A_2)$ is in blue.
	The blue sketch emphasizes only the ``heavy'' path $R\subseteq H_2$, while the red sketch shows all paths of the family $\ca Q$.
	The solid red paths are all of weight $w$ and the dashed red paths are of minimum weight $w-1$ (but some edges of these paths are also of weight~$w$).
	Note that every vertex of the graph $H_1$ is either a red anchor, or in the intersection of some two paths from~$\ca Q$.}
\label{fig:CM+setup}
\end{figure}

\begin{theorem2rep}[extension of {\Cref{thm:anchorNPH} \cite{DBLP:journals/siamcomp/CabelloM13}}]
\label{thm:anchorNPHrefin}\apxmark
Assume a PP anchored graph $(H,A)$, i.e., $H=H_1\cup H_2$ is a union of vertex-disjoint connected graphs,
and denoting by $A_i$, $i=1,2$, the restriction of $A$ to the anchors of $H_i$, the graphs $(H_1,A_1)$ and $(H_2,A_2)$ are anchored planar.
Let $\ca D_i$, $i=1,2$, be an anchored planar drawing of $(H_i,A_i)$.
Let $w$ be any sufficiently large integer parameter, which grows polynomially in the size of $H$.
Furthermore, assume the following (as informally illustrated in \Cref{fig:CM+setup}):
\begin{itemize}\parskip2pt
\item[a)] We have $A_1=(a_1,\ldots,a_{4k})$ where $k\in\mathbb N$, such that there are
\begin{itemize}
\item[--] for $i=1,\ldots,k-1$, a path $Q^1_i$ in $H_1$ from $a_i$ to $a_{3k+1-i}$ of weight $w-1$,
such that the edges of $Q^1_i$ incident to vertices in $A_1$ are of weight~$w$,
\item[--] a path $Q^1_k$ from $a_k$ to $a_{2k+1}$ of weight~$w$, and
\item[--] for $i=1,\ldots,k$, paths $Q^2_i$ and $Q^3_i$ in $H_1$ from $a_{i+k}$ to $a_{4k+1-i}$ of weight~$w$.
\end{itemize}
\vskip1pt
All paths $\ca Q=\{Q^1_i,Q^2_i,Q^3_i:i=1,\ldots,k\}$ are pairwise edge-dis\-joint, and
each of the unions $\bigcup_{i\in[k]}Q^1_i$ and $\bigcup_{i\in[k]}Q^2_i\cup Q^3_i$ spans~$V(H_1)\setminus A_1$.

\item[b)] We have $b,b'\in A_2$ such that $b$ is positioned between $a_1$ and $a_{4k}$ within the cyclic permutation~$A$,
and $b'$ is positioned between $a_{2k}$ and $a_{2k+1}$ within~$A$.
The graph $H_2$ contains a path $R\subseteq H_2$ of weight $w^{12}$ from $b$ to~$b'$.

\item[c)] All edges of $H_1$, and all edges of $H_2$ except those of $R$, have weight at most~$w^4$,
and the edges of $R$ in $H_2$ have weight at most $w^{12}+w^4$.

\item[d)] In every optimal solution to the anchored crossing number of $(H,A)$;
\begin{itemize}
\item[--] the subdrawing of $H_i$, $i\in\{1,2\}$, is homeomorphic to the drawing~$\ca D_i$,
\item[--] the path $R$ crosses the path $Q^1_i$, $1\leq i<k$, in an edge of weight $w-1$, and $R$ crosses the path $Q^1_k$
and the paths $Q^2_i$ and $Q^3_i$, $1\leq i\leq k$, in edges of weight~$w$.
\end{itemize}
\end{itemize}
Then it is \NP-hard to compute the anchored crossing number of~$(H,A)$.
\end{theorem2rep}

\begin{proof}
\begin{figure}[t]
$$\includegraphics[width=0.45\hsize]{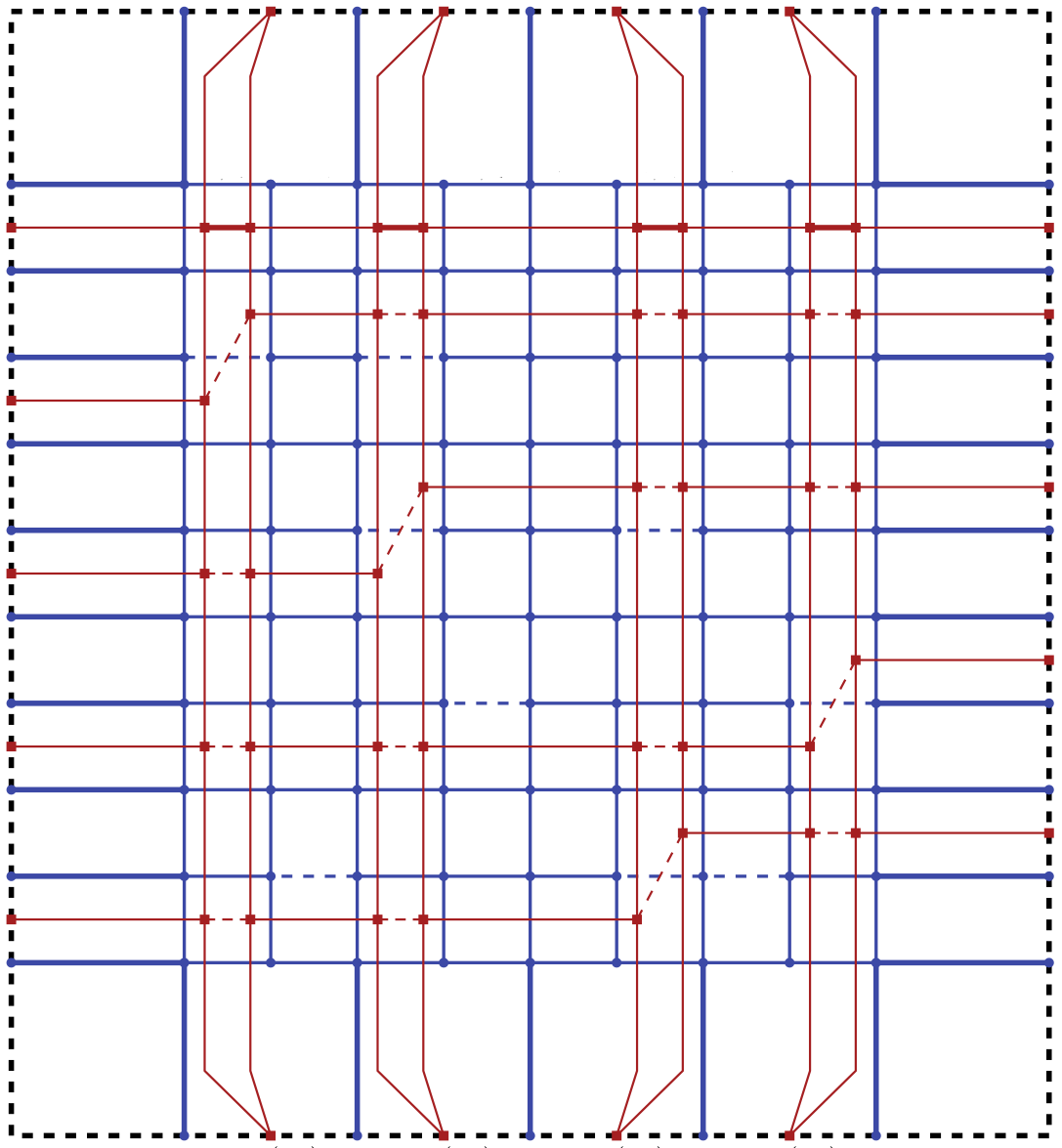}\qquad
\raise2ex\hbox{\vbox to 0.43\hsize{\hsize5ex\vfill$\begin{matrix}a_{2k+1}\\\\\\\vdots\\\\\\\vdots\\\\a_{3k}\end{matrix}$\vfill}
	\vbox{\hsize=0.4\hsize\hbox{$\qquad a_{2k}\qquad\cdots\quad\cdots\qquad a_{k+1}$}\smallskip
		\includegraphics[width=\hsize]{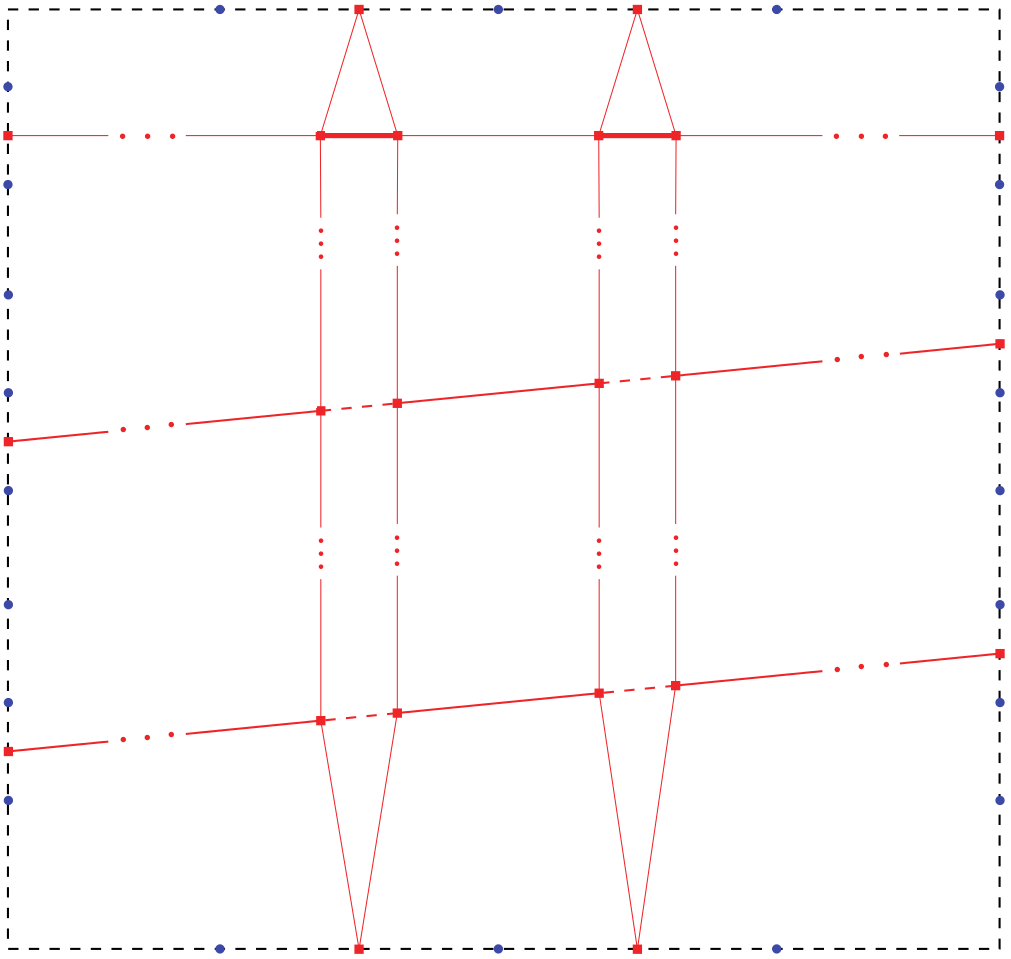}\\
	\hbox{$\qquad a_{3k+1}\quad\cdots\quad\cdots\qquad a_{4k}$}}
	\vbox to 0.44\hsize{\hsize2ex\vfill$\begin{matrix}a_k\\\\\vdots\\\\\\\\\vdots\\\\a_1\\\end{matrix}$\vfill}}
$$
\caption{An illustration of the starting step of the proof of Theorem~\ref{thm:anchorNPHrefin} --
	an example of the construction of a hard anchored crossing number instance $H_1\cup H_2$ from a SAT formula,
	copied from the arXiv:1203.5944 version of~\cite{DBLP:journals/siamcomp/CabelloM13}.
	Columns in the construction represent the variables of the formula, and rows represent the clauses.
	Here $H_2$ is blue, and $H_1$ is red (detailed alone with the anchors $A_1=(a_1,\ldots,a_{4k})$ on the right).
	The solid thin red edges all have weight $w$ (where $w$ is a large integer) and the middle dashed red edges have weight~$w-1$.
	For the blue solid and dashed thin edges, the weights are $w^2$ and $w^2-1$, respectively (with a slight exception
	of the topmost blue horizontal path), and the red and blue thick edges have weight $w^4$.}
\label{fig:CM}
\end{figure}
The reduction presented in \cite{DBLP:journals/siamcomp/CabelloM13}, as briefly sketched in \Cref{fig:CM},
reduces from a SAT formula $\varphi$ to a PP anchored instance $(H',A)$ where $H'=H'_1\cup H'_2$.
By possibly adding dummy positive-only variables, or by duplicating clauses, we may assume that the number $k$ of
variables in $\varphi$ is one more than the number of clauses in~$\varphi$.
Then the constructed instance satisfies all assumptions presented in Theorem~\ref{thm:anchorNPHrefin} except 
those concerning the ``very heavy'' path~$R$ in~$H_2$, as can be straightforwardly checked in \cite[Section~2.1]{DBLP:journals/siamcomp/CabelloM13}.
We thus let $H_1=H'_1$, and suitably add a path $R$ to $H'_2$ in order to produce~$H_2$.

For the latter task, we select a suitable ``diagonal'' path $R_2$ from to lower right corner of $H'_2$ (wrt.~\Cref{fig:CM} left) 
to the upper left corner, where some vertices of $R_2$ are subdividing some edges of $H'_2$, and increase weights of all edges of $R_2$ by $w^{12}$.
Formally, we turn $H'_2$ into $H''_2$ by subdividing some edges as specified below, select a sequence $R_2$ of vertices of $H''_2$, and define
$H_2:=H_2''\cup R$ where $R$ is a path following the vertex sequence $R_2$, but edge-disjoint from $H''_2$, and all edges of $R$ are of weight $w^{12}$.
By the Jordan curve theorem, in an anchored drawing, $R$ has to intersect each of the paths $Q^1_k$ and $Q^2_i,Q^3_i$ for $i=1,\ldots,k$ in an edge of weight~$w$,
and each of the paths $Q^1_i$ for $i=1,\ldots,k-1$ in an edge of weight at least $w-1$.
This construction will hence preserve the optimality of solutions, only numerically shifted up by $+(2k+1)w^{13}+(k-1)(w-1)w^{12}$,
if we can ensure by a choice of $R_2$ that the paths $Q^1_1,\ldots,Q^1_{k-1}$ indeed can cross $R$ in one of their weight-$(w-1)$ edges, 
which we discuss in the rest of the proof.

We explain, on a high level, the basic ideas of the reduction in \cite{DBLP:journals/siamcomp/CabelloM13}.
For every choice of a variable $x_i$, $1\leq i\leq k$, and a clause $c_j$, $1\leq j\leq k-1$, of the formula $\varphi$, 
the blue graph $H_2$ contains one $3\times3$ square subgrid $H_2[x_i,c_j]$ (i.e., four adjacent square blue faces in the picture).
Note that \Cref{fig:CM} indexes the variables right to left and the clauses bottom up.
Routing the red paths $Q^2_i,Q^3_i$ of $x_i$ through the left pair of square faces of $H_2[x_i,c_j]$ means selecting the value $x_i=T$.
Routing the red path $Q^1_j$ of the clause $c_j$ through the lower pair of square faces of $H_2[x_i,c_j]$ means that $c_j$ is not satisfied
by any of the variables $x_1,\ldots,x_i$, routing $Q^1_j$ through the upper pair of square faces of $H_2[x_i,c_j]$ means that $c_j$ is satisfied
by some of $x_1,\ldots,x_{i-1}$, and ``jumping up'' $Q^1_j$ from lower square(s) of $H_2[x_i,c_j]$ to upper square(s) means that $c_j$ is satisfied
by the literal of the variable~$x_i$.
The special path $Q^1_k$ has no such options, and causes no problems.

\begin{figure}[t]
{\centering\hfill a)
\begin{tikzpicture}[scale=1.2]\small
\tikzstyle{every node}=[draw, color=black, shape=circle, inner sep=1.1pt, fill=blue]
\tikzstyle{every path}=[draw, color=blue!80!black, ultra thick]
\draw (2,0)--(1,0) to[bend left] (0.5,1) --(0,2);
\tikzstyle{every path}=[draw, color=blue!80!black]
\draw (-0.2,0)--(2.2,0) (-0.2,1)--(2.2,1) (-0.2,2)--(2.2,2) ;
\draw (0,-0.2)--(0,2.2) (1,-0.2)--(1,2.2) (2,-0.2)--(2,2.2) ;
\draw node at (0,0) {} node at (0,1) {} node at (0,2) {};
\draw node at (1,0) {} node at (1,1) {} node at (1,2) {};
\draw node at (2,0) {} node at (2,1) {} node at (2,2) {};
\draw node at (0.5,1) {};
\tikzstyle{every node}=[draw, color=black, shape=circle, inner sep=1.0pt, fill=red!80!black]
\tikzstyle{every path}=[draw, color=red!70!black]
\draw (0.2,-0.3)--(0.2,2.3) (0.8,-0.3)--(0.8,2.3) (-0.3,0.5)-- (0.2,0.5) node {} (0.8,0.5) node {} --(2.3,0.5);
\draw[densely dashed] (0.2,0.5) -- (0.8,0.5);
\end{tikzpicture}
\hfill
\begin{tikzpicture}[scale=1.2]\small
\tikzstyle{every node}=[draw, color=black, shape=circle, inner sep=1.1pt, fill=blue]
\tikzstyle{every path}=[draw, color=blue!80!black, ultra thick]
\draw (2,0)--(1,0) -- (0.4,1) --(0,2);
\tikzstyle{every path}=[draw, color=blue!80!black]
\draw (-0.2,0)--(2.2,0) (-0.2,1)--(2.2,1) (-0.2,2)--(2.2,2) ;
\draw (0,-0.2)--(0,2.2) (1,-0.2)--(1,2.2) (2,-0.2)--(2,2.2) ;
\draw node at (0,0) {} node at (0,1) {} node at (0,2) {};
\draw node at (1,0) {} node at (1,1) {} node at (1,2) {};
\draw node at (2,0) {} node at (2,1) {} node at (2,2) {};
\draw node at (0.4,1) {};
\tikzstyle{every node}=[draw, color=black, shape=circle, inner sep=1.0pt, fill=red!80!black]
\tikzstyle{every path}=[draw, color=red!70!black]
\draw (0.2,-0.3)--(0.2,2.3) (0.8,-0.3)--(0.8,2.3) (-0.3,0.5)-- (0.2,0.5) node {} (0.8,1.5) node {} --(2.3,1.5);
\draw[densely dashed] (0.2,0.5) to[bend right=15] (0.8,1.5);
\end{tikzpicture}
\hfill
\begin{tikzpicture}[scale=1.2]\small
\tikzstyle{every node}=[draw, color=black, shape=circle, inner sep=1.1pt, fill=blue]
\tikzstyle{every path}=[draw, color=blue!80!black, ultra thick]
\draw (2,0)--(1,0) -- (0.5,1) to[bend right] (0,2);
\tikzstyle{every path}=[draw, color=blue!80!black]
\draw (-0.2,0)--(2.2,0) (-0.2,1)--(2.2,1) (-0.2,2)--(2.2,2) ;
\draw (0,-0.2)--(0,2.2) (1,-0.2)--(1,2.2) (2,-0.2)--(2,2.2) ;
\draw node at (0,0) {} node at (0,1) {} node at (0,2) {};
\draw node at (1,0) {} node at (1,1) {} node at (1,2) {};
\draw node at (2,0) {} node at (2,1) {} node at (2,2) {};
\draw node at (0.5,1) {};
\tikzstyle{every node}=[draw, color=black, shape=circle, inner sep=1.0pt, fill=red!80!black]
\tikzstyle{every path}=[draw, color=red!70!black]
\draw (0.2,-0.3)--(0.2,2.3) (0.8,-0.3)--(0.8,2.3) (-0.3,1.5)-- (0.2,1.5) node {} (0.8,1.5) node {} --(2.3,1.5);
\draw[densely dashed] (0.2,1.5) -- (0.8,1.5);
\end{tikzpicture}
\\[2ex]
\hfill b)
\begin{tikzpicture}[scale=1.2]\small
\tikzstyle{every node}=[draw, color=black, shape=circle, inner sep=1.1pt, fill=blue]
\tikzstyle{every path}=[draw, color=blue!80!black, ultra thick]
\draw (2,0)--(2,1) -- (1,1.4) to[out=170,in=290] (0.5,1.6) to[out=110,in=-30] (0,2);
\tikzstyle{every path}=[draw, color=blue!80!black]
\draw (-0.2,0)--(2.2,0) (-0.2,1)--(2.2,1) (-0.2,2)--(2.2,2) ;
\draw (0,-0.2)--(0,2.2) (1,-0.2)--(1,2.2) (2,-0.2)--(2,2.2) ;
\draw node at (0,0) {} node at (0,1) {} node at (0,2) {};
\draw node at (1,0) {} node at (1,1) {} node at (1,2) {};
\draw node at (2,0) {} node at (2,1) {} node at (2,2) {};
\draw node at (1,1.4) {};
\tikzstyle{every node}=[draw, color=black, shape=circle, inner sep=1.0pt, fill=red!80!black]
\tikzstyle{every path}=[draw, color=red!70!black]
\draw (0.2,-0.3)--(0.2,2.3) (0.8,-0.3)--(0.8,2.3) (-0.3,1.6)-- (0.2,1.6) node {} (0.8,1.6) node {} --(2.3,1.6);
\draw[densely dashed] (0.2,1.6) -- (0.8,1.6);
\end{tikzpicture}
\hfill
\begin{tikzpicture}[scale=1.2]\small
\tikzstyle{every node}=[draw, color=black, shape=circle, inner sep=1.1pt, fill=blue]
\tikzstyle{every path}=[draw, color=blue!80!black, ultra thick]
\draw (2,0)--(2,1) to[out=150,in=290] (1.5,1.4) to[out=110,in=-10] (1,1.6)--(0,2);
\tikzstyle{every path}=[draw, color=blue!80!black]
\draw (-0.2,0)--(2.2,0) (-0.2,1)--(2.2,1) (-0.2,2)--(2.2,2) ;
\draw (0,-0.2)--(0,2.2) (1,-0.2)--(1,2.2) (2,-0.2)--(2,2.2) ;
\draw node at (0,0) {} node at (0,1) {} node at (0,2) {};
\draw node at (1,0) {} node at (1,1) {} node at (1,2) {};
\draw node at (2,0) {} node at (2,1) {} node at (2,2) {};
\draw node at (1,1.6) {};
\tikzstyle{every node}=[draw, color=black, shape=circle, inner sep=1.0pt, fill=red!80!black]
\tikzstyle{every path}=[draw, color=red!70!black]
\draw (1.2,-0.3)--(1.2,2.3) (1.8,-0.3)--(1.8,2.3) (-0.3,1.4)-- (1.2,1.4) node {} (1.8,1.4) node {} --(2.3,1.4);
\draw[densely dashed] (1.2,1.4) -- (1.8,1.4);
\end{tikzpicture}
\hfill}
\caption{An addition to the proof of Theorem~\ref{thm:anchorNPHrefin}; how a section of the (very heavy) blue path $R$
	is inserted into the subgrid $H_2[x_i,c_i]$.
	~a) For odd indices $i$, the paths $Q^2_i$ and $Q^3_i$ are routed on the left side, and there are three possibilities
		for the path $Q^1_i$, as depicted -- in each one $R$ crosses $Q^1_i$ in a dashed edge of weight~$w-1$.
	~b) For even indices $i$, the path $Q^1_i$ is routed at the top, and there are two possibilites for the paths 
		$Q^2_i$ and $Q^3_i$, again as depicted -- in each one $R$ again crosses $Q^1_i$ in a dashed edge of weight~$w-1$.
}\label{fig:insertR}
\end{figure}

So, there are altogether $2\cdot3=6$ possible ways of routing the red edges across $H_2[x_i,c_j]$ in an optimal solution,
sketech above and proved in \cite{DBLP:journals/siamcomp/CabelloM13}.
Unfortunately, it is not possible to select an insertion of (a section of) $R$ in $H_2[x_i,c_j]$ to satisfy the desired properties in each of the $6$ ways of routing simultaneously.
To simply overcome this difficulty, we add a bunch of dummy all-positive variables in a bunch of dummy clauses satisfied by the first dummy variable $x_1$,
such that, overall, $x_1,x_3,\ldots,x_{k'}$ (every second) for $k'=2k+1$ are the dummy variables set to true $T$ in an optimal solution,
and that $c_2,c_4,\ldots,c_{k'-1}$ are the dummy clauses which take the upper routing of their paths $Q^1_j$ across $H_2[x_i,c_j]$ for~$i>1$.

Consider the sequence $H_2[x_i,c_i]$ of our subgrids for $i=1,2,\ldots,k'-1$.
For every $i$ odd, we know by the previous that the left routing of $Q^2_i$ and $Q^3_i$ can be taken, 
and for $i$ even the top routing of $Q^1_i$ can be taken, across $H_2[x_i,c_i]$.
We can thus insert the section of $R$ in $H_2[x_i,c_i]$ as defined in \Cref{fig:insertR} for all $i=1,2,\ldots,k'-1$,
and then simply finish $R$ across the topmost corridor of $Q^1_{k'}$ along edges of $H_2$,
which satisfies our demand.
\end{proof}

\section{Main Proof: the Hardness Reduction}\label{sec:hardness}

\subparagraph{Informal outline. }
Our proof of \Cref{thm:main3} is a bit complex, and we first informally explain what we want to achieve.
In a nutshell, we are going to ``embed'' the gadget $(H,A)$ of Theorem~\ref{thm:anchorNPHrefin} as an ordinary subgraph within
a special ``{\em frame}'' PP anchored graph $F_k$, parameterized by the gadget size,
such that $F_k$ forces the vertices of $A$ to be placed as expected in~$(H,A)$.
By the informal word ``forces'' we mean that any (other) drawing of $F_k$ violating the placement of the former anchors $A$ would
require increase in the number of crossings of $F_k$ higher than what is the difference between the best-case and the worst-case scenarios in Theorem~\ref{thm:anchorNPHrefin}.

We adopt the following ``colour coding''; the subgraph $H_1$ in Theorem~\ref{thm:anchorNPHrefin} will be called {\em red} and the subgraph $H_2$ {\em blue},
and these colours will be correspondingly used also within the frame~$F_k$ which will include some of the vertices and edges of $H$
(precisely, the special paths of $H$ claimed by Theorem~\ref{thm:anchorNPHrefin}(a,~b)).

\begin{figure}[t]
\hspace*{-5ex}
\begin{tikzpicture}[scale=2.6]\small
\draw[gray, ultra thick, densely dotted] (0,0) ellipse (80pt and 60pt);
\tikzstyle{every node}=[draw, color=black, thick, shape=circle, inner sep=1.3pt, fill=blue]
\node[label=below:$\!b_0\!$] (b0) at (-2.48,-1.0) {}; \node[label=below:$\!b_2\!$] (bM) at (0,-1.0) {};
\node[label=below:$b_4\!\!$,label=right:$~~P_1$] (bN) at (2.48,-1.0) {};
\node[label=below:$b_1\!\!$] (bL) at (-1.2,-1.0) {}; \node[label=below:$\!\!b_3$] (bR) at (1.2,-1.0) {};
\node[label=left:$c_0\!$] (bb0) at (-2.5,0.2) {}; 
\node[label=above:{\hspace*{-1.65ex}$c_2$\hspace*{1.5ex}}] (bbM) at (0,0.2) {};
\node[draw=none,fill=white] at (-0.12,0.75) {$R$};
\node[draw=none,fill=white] at (0.55,1.2) {$\ca Q$};
\node[label=right:$\!c_4\qquad P_2$] (bbN) at (2.5,0.2) {};
\node[label=above:$c_1\!\!$] (bbL) at (-1.2,0.2) {}; \node[label=above:$c_3\!\!$] (bbR) at (1.2,0.2) {};
\node[label=above:$d_0~$] (bbbL) at (-1.02,1.9) {}; \node (bbbM) at (0,1.9) {};
\node[label=above:$~d_4$] (bbbR) at (1.02,1.9) {}; 
\node[label=below:{\,~$d_2$\,~~~~~~}, label=above:$d_1$] (bT) at (0,2.10) {};
\tikzstyle{every node}=[draw, color=black!50!blue, shape=circle, inner sep=0.85pt, fill=blue]
\node (b1) at (-2.2,-1.0) {}; \node (b2) at (-1.95,-1.0) {}; \node (b3) at (-1.7,-1.0) {}; \node (b4) at (-1.45,-1.0) {};
\node (bb2) at (-1.95,0.2) {}; \node (bb3) at (-1.7,0.2) {}; \node (bb4) at (-1.45,0.2) {};
\node (b5) at (-0.95,-1.0) {}; \node (b6) at (-0.7,-1.0) {}; \node (b7) at (-0.45,-1.0) {}; \node (b8) at (-0.2,-1.0) {};
\node (bb5) at (-0.95,0.2) {}; \node (bb6) at (-0.7,0.2) {}; \node (bb7) at (-0.45,0.2) {}; \node (bb8) at (-0.2,0.2) {};
\node (b1') at (2.2,-1.0) {}; \node (b2') at (1.95,-1.0) {}; \node (b3') at (1.7,-1.0) {}; \node (b4') at (1.45,-1.0) {};
\node (bb2') at (1.95,0.2) {}; \node (bb3') at (1.7,0.2) {}; \node (bb4') at (1.45,0.2) {};
\node (b5') at (0.95,-1.0) {}; \node (b6') at (0.7,-1.0) {}; \node (b7') at (0.45,-1.0) {}; \node (b8') at (0.2,-1.0) {};
\node (bb5') at (0.95,0.2) {}; \node (bb6') at (0.7,0.2) {}; \node (bb7') at (0.45,0.2) {}; \node (bb8') at (0.2,0.2) {};
\tikzstyle{every path}=[draw, color=blue!80!black, ultra thick]
\draw (b0)--(bM)--(bN) (bbbL)--(bbbM)--(bbbR);
\draw (bM)--(bbM) (bbbM)--(bT);
\draw[dash pattern=on 4pt off 1pt] (bbM)--(bbbM);
\draw (bb0) to[bend left=36] (bT);
\draw (bT) to[bend left=36] (bbN);
\tikzstyle{every path}=[draw, color=blue, very thick]
\draw (bb0)--(bbM)--(bbN) (bL)--(bbL) (bR)--(bbR);
\tikzstyle{every path}=[draw, color=blue, thick]
\draw (b1) to[bend right=19] (bb0) (b2)--(bb2) (b3)--(bb3) (b4) to[bend left=6] (bb4);
\draw (b5)--(bb5) (b6)--(bb6) (b7)--(bb7) (b8)--(bb8);
\draw (b1') to[bend left=19] (bbN) (b2')--(bb2') (b3')--(bb3') (b4') to[bend right=6] (bb4');
\draw (b5')--(bb5') (b6')--(bb6') (b7')--(bb7') (b8') to[bend right=12] (bb8');

\tikzstyle{every node}=[draw, color=black, thick, shape=circle, inner sep=1.2pt, fill=red!80!black]
\node[label=left:$r_0\!$] (r0) at (-2.76,-0.4) {}; \node[label=right:$\!r_4~P_0$] (rN) at (2.76,-0.4) {};
\node[label=above:$r_1\!$] (rL) at (-0.13,-0.4) {}; \node[label=above:$r_3\!$] (rR) at (0.13,-0.4) {};
\node[label=below:$r_2$] (rB) at (0,-2.11) {};  \node[draw=none,fill=white] at (0,-1.78) {$C\!_0$};
\node (rrL) at (-0.13,-1.5) {}; \node (rrR) at (0.13,-1.5) {};
\tikzstyle{every node}=[draw, color=black!50!red, shape=circle, inner sep=0.85pt, fill=red!90!black]
\node (r1) at (-2.07,-0.4) {}; \node (r2) at (-1.82,-0.4) {}; \node (r3) at (-1.57,-0.4) {}; \node (r4) at (-1.32,-0.4) {};
\node (r5) at (-1.07,-0.4) {}; \node (r6) at (-0.82,-0.4) {}; \node (r7) at (-0.57,-0.4) {}; \node (r8) at (-0.32,-0.4) {};
\node (r1') at (2.07,-0.4) {}; \node (r2') at (1.82,-0.4) {}; \node (r3') at (1.57,-0.4) {}; \node (r4') at (1.32,-0.4) {};
\node (r5') at (1.07,-0.4) {}; \node (r6') at (0.82,-0.4) {}; \node (r7') at (0.57,-0.4) {}; \node (r8') at (0.32,-0.4) {};
\tikzstyle{every path}=[draw, color=red!80!black, ultra thick]
\draw (r0)--(rL)--(rR)--(rN) (rrL)--(rrR);
\draw (rB) to[bend left=9] (rrL) (rrR) to[bend left=9] (rB) (rrL)--(rL) (rrR)--(rR);
\tikzstyle{every path}=[draw, color=red!90!black, thick]
\tikzstyle{every node}=[draw, color=black!40!red, shape=circle, inner sep=0.6pt, fill=red!80!black]
\draw (r1)-- (-2.07,-1.0) to[out=277,in=175] (-0.04,-2) node {};
\draw (r2)-- (-1.82,-1.0) to[out=275,in=175] (-0.055,-1.95) node {};
\draw (r3)-- (-1.57,-1.0) to[out=275,in=175] (-0.068,-1.90) node {};
\draw (r4)-- (-1.32,-1.0) to[out=275,in=175] (-0.08,-1.85) node {};
\draw (r5)-- (-1.07,-1.0) to[out=273,in=175] (-0.09,-1.80) node {};
\draw (r6)-- (-0.82,-1.0) to[out=272,in=175] (-0.10,-1.75) node {};
\draw (r7)-- (-0.57,-1.0) to[out=270,in=175] (-0.11,-1.70) node {};
\draw (r8)-- (-0.32,-1.0) to[out=270,in=175] (-0.118,-1.65) node {};
\draw (r1')-- (2.07,-1.0) to[out=263,in=5] (0.04,-2) node {};
\draw (r2')-- (1.82,-1.0) to[out=265,in=5] (0.055,-1.95) node {};
\draw (r3')-- (1.57,-1.0) to[out=265,in=5] (0.068,-1.90) node {};
\draw (r4')-- (1.32,-1.0) to[out=265,in=5] (0.08,-1.85) node {};
\draw (r5')-- (1.07,-1.0) to[out=267,in=5] (0.09,-1.80) node {};
\draw (r6')-- (0.82,-1.0) to[out=268,in=5] (0.10,-1.75) node {};
\draw (r7')-- (0.57,-1.0) to[out=270,in=5] (0.11,-1.70) node {};
\draw (r8')-- (0.32,-1.0) to[out=270,in=5] (0.118,-1.65) node {};
\tikzstyle{every path}=[draw, color=red!90!black, dash pattern=on 4pt off 1pt]
\draw (r5)-- (-1.07,0.3) to[out=90,in=180] (0.3,1.70) to[out=0,in=90] (2.07,0.3) --(r1');
\draw (r5)-- (-1.09,0.3) to[out=90,in=180] (0.3,1.73) to[out=0,in=90] (2.09,0.3) --(r1');
\draw (r6)-- (-0.82,0.4) to[out=90,in=180] (0.3,1.50) to[out=0,in=90] (1.82,0.3) --(r2');
\draw (r6)-- (-0.84,0.4) to[out=90,in=180] (0.3,1.53) to[out=0,in=90] (1.84,0.3) --(r2');
\draw (r7)-- (-0.57,0.4) to[out=90,in=180] (0.3,1.30) to[out=0,in=90] (1.57,0.3) --(r3');
\draw (r7)-- (-0.59,0.4) to[out=90,in=180] (0.3,1.33) to[out=0,in=90] (1.59,0.3) --(r3');
\draw (r8)-- (-0.32,0.5) to[out=90,in=180] (0.3,1.10) to[out=0,in=90] (1.32,0.3) --(r4');
\draw (r8)-- (-0.34,0.5) to[out=90,in=180] (0.3,1.13) to[out=0,in=90] (1.34,0.3) --(r4');
\draw (r1)-- (-2.07,0.3) to[out=90,in=180] (-0.4,1.70) to[out=0,in=90] (1.07,0.3) --(r5');
\draw (r2)-- (-1.82,0.4) to[out=90,in=180] (-0.4,1.50) to[out=0,in=90] (0.82,0.3) --(r6');
\draw (r3)-- (-1.57,0.4) to[out=90,in=180] (-0.4,1.30) to[out=0,in=90] (0.57,0.3) --(r7');
\draw (r4)-- (-1.32,0.5) to[out=90,in=180] (-0.4,1.10) to[out=0,in=90] (0.32,0.3) --(r8');

\def\eweight#1{{\scriptsize$[#1]$}}
\tikzstyle{every node}=[rectangle, inner sep=0.4pt, draw=none, color=blue!80!black, fill=white]
\node at (0.15,2) {\eweight{49}}; \node at (-0.15,2) {\eweight{49}};
\node at (2.5,0.6) {\eweight{49}}; \node at (-2.5,0.6) {\eweight{49}};
\node at (0.12,-0.06) {\eweight{48}}; \node at (0.12,0.5) {\eweight{48}};
\node at (1.32,-0.2) {\eweight{35}}; \node at (-1.32,-0.2) {\eweight{35}};
\node at (0.58,-0.2) {\eweight{30}}; \node at (-0.58,-0.2) {\eweight{30}};
\node at (2.13,-0.2) {\eweight{30}}; \node at (-2.13,-0.2) {\eweight{30}};
\node at (2.4,-0.9) {\eweight{49}}; \node at (-2.4,-0.9) {\eweight{49}};
\node at (2.25,0.3) {\eweight{38}}; \node at (-2.25,0.3) {\eweight{38}};
\node[color=black, fill=white] at (-0.3,-1.5) {$r_1'\!$};
\node[color=black, fill=white] at (0.3,-1.5) {$r_3'$};
\tikzstyle{every node}=[rectangle, inner sep=0pt, draw=none, color=red!70!black, fill=white]
\node at (2.55,-0.31) {\eweight{41}}; \node at (-2.55,-0.31) {\eweight{41}};
\node at (0,-1.6) {\eweight{49}}; \node at (0,-1.35) {\eweight{41}};
\node at (0.44,-1.2) {\eweight{30}}; \node at (-0.44,-1.2) {\eweight{30}};
\node at (1.9,-1.2) {\eweight{30}}; \node at (-1.9,-1.2) {\eweight{30}};
\node at (2.15,0.6) {\eweight{4}}; \node at (-2.12,0.6) {\eweight{4}};
\node at (0.45,0.5) {\eweight{4}}; \node at (-0.45,0.5) {\eweight{4}};

\node[color=black, fill=none] at (-2.4,1.8) {\small\it gadget region};
\draw[gray, thick, dashed, ->] (-2.2,1.7)--(-1.2,1.4);
\begin{scope}[on background layer]
\draw[black!15!white, solid, line width=2mm] (-2.75,-0.4)--(2.75,-0.4);
\draw[black!15!white, solid, line width=2mm] (-2.5,-1)--(2.5,-1);
\draw[black!15!white, solid, line width=1.7mm] (0,0.2)--(0,-1);
\draw[black!15!white, solid, line width=1.7mm] (0.13,-0.4)--(0.13,-1.5);
\draw[black!15!white, solid, line width=1.7mm] (-0.13,-0.4)--(-0.13,-1.5);
\draw[black!15!white, solid, line width=2mm] (1.2,0.2)--(2.5,0.2);
\draw[black!15!white, solid, line width=2mm] (-1.2,0.2)--(-2.5,0.2);
\end{scope}
\end{tikzpicture}

\caption{A schematic picture of the frame PP anchored graph $(F_k,B)$ used for the reduction in \Cref{sec:hardness} with $k=4$.
	The solid lines depict edges, while the dashed lines represent paths in general.
	The thin dashed red paths pairwise intesect in red vertices which are not (and do not need to be) explicitly specified.
	The red and the blue graphs are vertex-disjoint and each one has three anchors 
	($r_0,r_2,r_4$ of the red graph, and $b_0,d_1,b_4$ of the blue graph) and is anchored planar.
	The bracketed numbers represent edge weights, where $[t]$ means weight $\omega^t$.
	\\Weights of the edges emphasized with gray shade are treated specially (see a closer detail in \Cref{fig:Fk-order}).
	The horizontal red path $P_0$ from $r_0$ to $r_4$ has edges of weight $\omega^{41}+\ca O(k\omega^{30})$,
	the horizontal blue path $P_1$ connecting $b_0$ to $b_4$ has, in the subpath from $b_1$ to $b_3$, edges of weight $\omega^{49}+\ca O(k\omega^{30})$,
	and elsewhere edges of weight $\omega^{49}+\frac25\omega^{35}+\ca O(k\omega^{30})$;
	these weights are precisely specified later in the proof.
	The horizontal blue path $P_2$ connecting $c_0$ to $c_4$ has edges of weight exactly $\omega^{38}$ in the subpath
	from $c_1$ to $c_3$, and of weight $\omega^{38}+\frac45\omega^{35}$ elsewhere.
	The vertical blue edge $b_2c_2$ is of weight exactly $\omega^{48}+2\omega^{38}-\omega^{34}$,
	and the vertical red edges $r_1r_1'$ and $r_3r_3'$ are of weight $\omega^{41}-\omega^{40}$.
}\label{fig:Fk-all}
\end{figure}

For the sole purpose of this informal outline, it is enough to ``define'' the frame PP anchored graph $(F_k,B)$ via a detailed sketch in \Cref{fig:Fk-all}.
The key feature of $(F_k,B)$ is the use of ``heavy-weight'' edges, whose weight is of the form $\omega^t$ where $t$ is specified at each edge
in the picture and $\omega$ is now seen as a variable base.
Later, $\omega$ is chosen as a ``sufficiently large'' integer such that for every $t$, 
$\>\omega^{t+1}$ is always larger than the sum of a collection of crossings of weight at most $\omega^t$ in the reduction
(with a slight abuse of traditional notation, $\omega^{t+1}>\Theta(\omega^t)$), and that all weights are integers.

Briefly, the blue graph of $F_k$ has $3$ anchors $B_2=\{b_0,b_4,d_1\}\subseteq B$, and consists of the vertices lying on two horizontal paths 
$P_1$ from $b_0$ to $b_4$ of length $4k+4$ and $P_2$ from $c_0$ to $c_4$ of length $4k+2$,
four special vertices $d_0,d_1,d_2,d_4$, and the vertices of a vertical path $R$ from $c_2$ to~$d_2$ of weight exactly $\omega^{48}$
(cf.\ Theorem~\ref{thm:anchorNPHrefin}(b) with $w=\omega^4$).
Additional edges exist in $F_k$ between the specified blue vertices as sketched in \Cref{fig:Fk-all} and detailed in the full definition.
The red graph of $F_k$ has again $3$ anchors $B_1=\{r_0,r_2,r_4\}\subseteq B$, and consists of the vertices lying on a cycle $C_0$ of length $4k+3$ passing through $r_2$,
the vertices on a horizontal path $P_0$ from $r_0$ to $r_4$ of length $4k+3$, and (the vertices of) a collection $\ca Q$ of $k+2k=3k$ edge-disjoint paths
of weight $\omega^4-1$ or $\omega^4$ (as in Theorem~\ref{thm:anchorNPHrefin}(a) with $w=\omega^4$) which connect pairs of vertices of $P_0$.
Again, further edges exist in $F_k$ between vertices of the red path $P_0$ and the red cycle $C_0$, as sketched in \Cref{fig:Fk-all} and detailed later.
Moreover, all end-edges of the paths in~$\ca Q$ (i.e., those incident to $P_0$) are of weight $\omega^4$,
and the second condition of Theorem~\ref{thm:anchorNPHrefin}(d) is met by the paths in~$\ca Q$.

Note that we do not require to exactly specify the paths of $\ca Q$, in particular they share their internal vertices in an unspecified way
(and so, we rather construct a family of frame graphs $F_k$ for each~$k$ than single~$F_k$),
but we do require the properties stated in Theorem~\ref{thm:anchorNPHrefin} to hold for them.
With respect to the drawing in \Cref{fig:Fk-all}, we call the {\em gadget region} of $(F_k,B)$ the region bounded by the blue horizontal path
from $c_0$ to $c_4$ and the path~$(c_0,d_0,d_2,d_4,c_4)$.
Summarizing, our coming proof proceeds in the following points:
\begin{itemize}
\item (\Cref{lem:framecr})
Let $\gamma_\omega(k)$ denote the number of crossings in the drawing of $(F_k,B)$ as in \Cref{fig:Fk-all} (see \Cref{lem:framecr} for an exact formula).
In the setting of weighted edges (cf.~\Cref{sec:prelim}), we claim roughly the following;
any drawing of $(F_k,B)$ which is ``different'' from the one in \Cref{fig:Fk-all} or, specially, in which not all internal vertices of the
edges of~$\ca Q$ lie in the gadget region, has at least $\gamma_\omega(k)+\omega^{34}-\omega^{30}$ crossings.
\item (\Cref{thm:main3})
Based on the previous point, we can use the internal vertices of the red path $P_0$ and of the blue path $P_2$ and the vertex~$d_2$
as ``firmly glued'' emulated anchors $A$ for the gadget anchored graph $(H,A)$ of Theorem~\ref{thm:anchorNPHrefin}\,---\,if these emulated anchors $A$
were not placed as required, then the increase in the number of crossings ($\omega^{34}-\omega^{30}$) of the frame $(F_k,B)$ would be strictly larger than
the number of crossings used to draw $(H,A)$ itself (on top of the crossings existing in the subdrawing of $(F_k,B)$).
We thus reduce the problem of determining the anchored crossing number of $(H,A)$ to that of $(F_k\cup H,B)$.
\end{itemize}

\subparagraph{The ``frame'' graph in detail. }
As previously sketched in \Cref{fig:Fk-all}, a PP anchored graph $(F_k,B)$ is called a {\em frame graph} of the parameter~$k$ and weight~$\omega$
if the following holds:
$(F_k,B)$ is constructed as a disjoint union $F_k=F_k^1\cup F_k^2$ and $B=B_1\cup B_2$,
where the component $F_k^1$ is coded as \emph{red} and $F_k^2$ as \emph{blue}.
In red $F_k^1$, we have the anchors $B_1=\{r_0,r_2,r_4\}$, a cycle $C_0$ passing through $r_2$, a path $P_0$ from $r_0$ to $r_4$, and:
\begin{itemize}
\item
The vertices of $P_0$ are in order $V(P_0)=(r_0,r_0^1,\ldots,r_0^{2k},r_1,r_3,r_0^{2k+1},\ldots,r_0^{4k},r_4)$.
The weights of its edges are $\omega^{41}+\ca O(k\omega^{30})$ and are exactly specified below in the proof of \cref{lem:framecr}.
A collection of (red) paths $\ca Q$ satisfies the assumptions of Theorem~\ref{thm:anchorNPHrefin}, their ends are identified with vertices of $P_0$
as $a_i=r_0^i$ for $i=1,\ldots,4k$, and their weights are set according to (the gadget $(H,A)$ of) Theorem~\ref{thm:anchorNPHrefin} and $w:=\omega^4$.
\item
The vertices of $C_0$ are in cyclic order $V(C_0)=(r_2,s_0^1,\ldots,s_0^{2k},r_1',r_3',s_0^{2k+1},\ldots,$ $s_0^{4k},r_2)$.
The edges of $C_0$ are all of weight $\omega^{49}$, and there are additional two edges $r_1'r_1$ and $r_3'r_3$ of weight $\omega^{41}-\omega^{40}$
and $4k$ edges $r_0^is_0^i$ of weight $\omega^{30}$ for $i=1,\ldots,4k$.
\end{itemize}
In blue $F_k^2$, we have the anchors $B_2=\{b_0,d_1,b_4\}$, a path $P_1$ from $b_0$ to $b_4$, a path $P_2$ from $c_0$ to $c_4$,
a path $R$ from $c_2$ to $d_2$, three vertices $d_0,d_1,d_4$, and:
\begin{itemize}
\item
The vertices of $P_1$ are in order $V(P_1)=(b_0,b_0^1,\ldots,b_0^{k},b_1,b_0^{k+1},\ldots,b_0^{2k},b_2,b_0^{2k+1},\ldots,b_0^{3k},$ $b_3,b_0^{3k+1},\ldots,b_0^{4k},b_4)$.
The vertices of $P_2$ are in order $V(P_2)=(c_0=c_0^1,\ldots,c_0^{k},c_1,$ $c_0^{k+1},\ldots,c_0^{2k},c_2,c_0^{2k+1},\ldots,c_0^{3k},c_3,c_0^{3k+1},\ldots,c_0^{4k}=c_4)$.
The edges of $P_1$ are of weight $\omega^{49}+\ca O(k\omega^{30})$ between $b_1$ and $b_3$ and of weight $\omega^{49}+\ca O(k\omega^{34})+\ca O(k\omega^{30})$
elsewhere, and are exactly specified below in the proof.
The edges of $P_2$ are of weight $\omega^{38}$ between $c_1$ and $c_3$ and $\omega^{38}+\frac45\omega^{35}$ elsewhere.
\item
There is an edge $b_2c_2$ of weight $\omega^{48}+2\omega^{38}-\omega^{34}$, two edges
$b_1c_1$ and $b_3c_3$ of weight $\omega^{35}$, and $4k$ edges $b_0^ic_0^i$ of weight $\omega^{30}$ for $i=1,\ldots,4k$.
\item
There is a path $R$, as specified in Theorem~\ref{thm:anchorNPHrefin}(b), from $c_2$ to $d_2$ of weight~$\omega^{48}$.
There are seven additional edges $c_0d_0$, $d_0d_1$, $d_0d_2$, $d_1d_4$, $d_2d_4$, $d_4c_4$, each of weight~$\omega^{49}$.
\end{itemize}
One can easily check from \Cref{fig:Fk-all} that each of $(F_k^1,B_1)$ and $(F_k^2,B_2)$ is anchored planar.

\medskip
We have the following claim.

\begin{lemma2rep}\label{lem:framecr}\apxmark
Let a PP anchored graph $(F_k,B)$ be a frame graph for $k$ and $\omega$.
For $k\geq2$ and any sufficiently large $\omega$ (divisible by $5(5k+7)$ to maintain integrality), its anchored crossing number equals
\begin{eqnarray*}
\cra(F_k,B)=\gamma_\omega(k) :=&& 2\omega^{90}-\omega^{89}+ (4k+2)\omega^{79} + 2\omega^{76}-\omega^{75}+4k\omega^{71}
\\
	&+& c_1(k)\omega^{65}+c_2(k)\omega^{60} + 3k\omega^{52}-k\omega^{48}+6k\omega^{42} +\frac{12}5k\omega^{39}
,\end{eqnarray*}
where $c_1(k)=\frac2{5(5k+7)}(30k^2+58k+20)$ and $c_2(k)=\frac2{3(5k+7)}(16k^3+39k^2+20k)$.

\smallskip
Any anchored drawing of $(F_k,B)$ with at most $\gamma_\omega^+(k):=\gamma_\omega(k)+\omega^{34}-\omega^{30}-1$ crossings
is homeomorphic, with a possible exception of the paths of $\ca Q$, to that in \Cref{fig:Fk-all}, and specially the internal vertices 
of all paths of $\ca Q$ are drawn in the gadget region and their edges cross the path $P_2$ from $c_0$ to $c_4$ and the path $R$ as depicted.
\end{lemma2rep}

\ifx\proof\inlineproof\else
\begin{proof}
	{$\!\!$\sf\bfseries\color{lipicsGray}(Step \ref{it:rbordering})\,}
We repeat the proof of step \eqref{it:rbordering} in the proof of \Cref{lem:framecr} with full details.

\medskip
We first note that the graph $(F_k,B)$ is symmetric along the vertical axis, except the paths of~$\ca Q$.
Since the paths of $\ca Q$ are handled only after point \eqref{it:rbordering}, we can now assume full symmetry and prove the claim only for the right-hand side
of $(F_k,B)$, as depicted in \Cref{fig:Fk-order}, and then multiply the number of crossings by~$2$.

We recapitulate where we stand with our drawing of $(F_k,B)$ after \eqref{it:topbound}--\eqref{it:thirdx}.
We have got pairwise noncrossing cycle $C_0$ and paths $P_1$, $P_0$ and $P_2$ drawn in this order bottom up (\Cref{fig:Fk-all}).
All crossings of weight of order $\omega^{66}$ and higher have already been counted to equality with~$\gamma_\omega(k)$.
There are two edges of weight $\omega^{41}-\omega^{40}$ between $C_0$ and $P_0$ crossing $P_1$, and $4k$ such edges of weight $\omega^{30}$.
There is an edge of weight $\omega^{48}+2\omega^{38}-\omega^{34}$ and two edges of weight $\omega^{35}$ between $P_1$ and $P_2$ crossing $P_0$,
and again $4k$ such edges of weight $\omega^{30}$.
Besides the already counted weights, these listed edges contribute crossings of weights of order only $\omega^{65}$ and $\omega^{60}$
(including their possible mutual crossings).

Our proof strategy is the following; we will show a concrete drawing, called the \emph{normal drawing}, in which the above described crossings contribute exactly as expected 
in the formula for~$\gamma_\omega(k)$ in the part $c_1(k)\omega^{65}+c_2(k)\omega^{60}$ (the drawing in \Cref{fig:Fk-order}),
and then we will argue that in any drawing in which the above listed edges (the vertical blue and red ones of weights $\omega^{35}$and $\omega^{30}$)
are not as in our normal drawing, we can decrease the total crossing number by $\Omega(\omega^{60})$.
This possible drop in the crossing number in turn certifies that the arbitrary considered drawing would exceed $\gamma_\omega^+(k)$ crossings,
and hence is impossible in our claim.
In this setup of a proof, it also comes for free that all crossings potentially occuring in $(F_k,B)$, 
which are not counted prior to this point and are not among the crossings listed above, are of weight of order strictly less than~$\omega^{60}$.

The weights of the edges of $P_0$ and $P_1$ are precisely as follows (\Cref{fig:Fk-order}):
\begin{itemize}\parskip2pt
\item For $P_0$, $V(P_0)=(r_0,r_0^1,\ldots,r_0^{2k},r_1,r_3,r_0^{2k+1},\ldots,r_0^{4k},r_4)$,
\smallskip
the weight of $r_1r_3$ is exactly $\omega^{41}$, the weight of $r_0^{2k}r_1$ and of $r_3r_0^{2k+1}$ is $\omega^{41}\!+\frac{2}{5k+7}\omega^{30}$,
the weight of $r_0^{2k+i-1}r_0^{2k+i}$ and of $r_0^{2k-i+1}r_0^{2k-i}$ is $\omega^{41}\!+\frac{i(i+1)}{5k+7}\omega^{30}$ for $i=2,\ldots,2k$,
and the weight of $r_0r_0^1$ and of $r_0^{4k}r_4$ is $\omega^{41}\!+\frac{(2k+1)(2k+2)}{5k+7}\omega^{30}$.

\item For $P_1$, $V(P_1)=(b_0,b_0^1,\ldots,b_0^{k},b_1,b_0^{k+1},\ldots,b_0^{2k},b_2,b_0^{2k+1},\ldots,b_0^{3k},b_3,b_0^{3k+1},$ $\ldots,b_0^{4k},b_4)$,
we resort to describing the weights only from $b_2$ till $b_4$.
The weight of $b_2b_0^{2k+1}$ is exactly $\omega^{49}$, 
the weight of $b_0^{2k+i}b_0^{2k+i+1}$ is $\omega^{49}\!+\frac{i(i+2)}{5k+7}\omega^{30}$ for $i=1,\ldots,k-1$,
the weight of $b_0^{3k}b_3$ is $\omega^{49}\!+\frac{k(k+2)}{5k+7}\omega^{30}$, 
the weight of $b_3b_0^{3k+1}$ is $\omega^{49}\!+\frac25\omega^{35}\!+\frac{(k+1)(k+3)}{5k+7}\omega^{30}$,
the weight of $b_0^{3k+i}b_0^{3k+i+1}$ is $\omega^{49}\!+\frac25\omega^{35}\!+\frac{(i+k+1)(i+k+3)}{5k+7}\omega^{30}$ for $i=1,\ldots,k-1$,
and the weight of $b_0^{4k}b_4$ is $\omega^{49}\!+\frac25\omega^{35}\!+\frac{(2k+1)(2k+3)}{5k+7}\omega^{30}$.
\end{itemize}

We count the crossings in a normal drawing as in \Cref{fig:Fk-order}:
\begin{itemize}
\item Concerning total crossings of weight of order $\omega^{65}$, 
we have $2k$ red edges of weight $\omega^{30}$ crossing the sections of $P_1$ between $b_0$--$b_1$ and $b_3$--$b_4$,
and two blue edges of weight $\omega^{35}$ crossing $P_0$ in edges of weight $\frac{(k+1)(k+2)}{5k+7}$. In total 
\begin{eqnarray*}
2k\omega^{30}\cdot\frac25\omega^{35}+2\omega^{35}\cdot\frac{(k+1)(k+2)}{5k+7}\omega^{30}=\frac2{5(5k+7)}(30k^2+58k+20)\omega^{65}
 =c_1(k)\omega^{65}.
\end{eqnarray*}
\item Concerning total crossings of weight of order $\omega^{60}$, and considering only the right-hand side as in \Cref{fig:Fk-order},
we get crossings of $2k$ red edges of weight $\omega^{30}$ with the edges of $P_1$ from $b_0^{2k+1}$ to $b_0^{4k}$,
and crossings of $2k$ blue edges of weight $\omega^{30}$ with the edges of $P_0$ from $r_3$ to $r_4$ except with the edge of weight $\frac{(k+1)(k+2)}{5k+7}$
(counted in the previous point). These sum to
\begin{eqnarray*}
\omega^{30}\cdot&&\sum_{i=1}^{2k}\frac{i(i+2)}{5k+7}\omega^{30} + \omega^{30}\cdot\left[\,\sum_{i=1}^{2k+1}\frac{i(i+1)}{5k+7}-\frac{(k+1)(k+2)}{5k+7}\right]\omega^{30}
\\&&=\frac1{3(5k+7)}(16k^3+39k^2+20k)
,\end{eqnarray*}
which multiplied by $2$ gives~$c_2(k)$.
\end{itemize}

Now, we handle an arbitrary drawing of $(F_k,B)$ which conforms to the bounds shown in \eqref{it:topbound}--\eqref{it:thirdx}, 
but is not our normal drawing described by \Cref{fig:Fk-order}.
As argued above, it is enough for this purpose to consider only edges depicted in \Cref{fig:Fk-order}, and know that the depicted vertical
edges indeed cross the horizontal paths $P_1$ and $P_2$ somewhere. 
Crossings of weight of order strictly higher than $\omega^{65}$ are not relevant in this analysis (as they are enforced and have been counted,
and so new such ones cannot even arise), and crossings of weight of order strictly less than $\omega^{60}$ can be ignored at this stage (by the choice of sufficiently large $\omega$).

We denote by $t^0_i=\frac{i(i+1)}{5k+7}$ and $t^1_i=\frac{i(i+2)}{5k+7}$, and refer to \Cref{fig:Fk-order}.
The $i$-th edge of the red path $P_0$, counted from $r_3$ towards $r_4$, is of weight $\omega^{41}\!+t^0_i\omega^{30}$,
and the $(i+1)$-th edge of the blue path $P_1$, counted from $b_2$ towards $b_4$, is of weight $\omega^{49}\!+t^1_i\omega^{30}$,
resp.~of weight $\omega^{49}\!+\frac25\omega^{35}\!+t^1_i\omega^{30}$ if~$i>k$.
We also denote the vertical red edges (of weight $\omega^{30}$) incident to the path $P_0$ by $f^0_1,\ldots,f^0_{2k}$ such that
$f^0_i$ is incident to the vertex $r_0^{2k+i}$, and the vertical blue edges (of weight again~$\omega^{30}$) incident to the path $P_1$ 
by $f^1_1,\ldots,f^1_{2k+1}$ such that $f^1_i$ is incident to the vertex $b_0^{2k+i}$ for $i\leq k$,~ 
$f^1_{k+1}$ is incident to $b_3$, and $f^1_i$ is incident to $b_0^{2k+i-1}$ for $i\geq k+2$.

One can easily check that in our normal drawing, we have, naturally, $f^0_i$ crossing $P_1$ in the edge of weight coefficient $t^1_i$,
~$f^1_j$ crossing $P_0$ in the edge of weight coefficient $t^0_j$, and no $f^0_i$ is crossing any $f^1_j$.
Assume that the drawing of $f^0_i$ violates some of the previous conditions.
\begin{enumerate}[i.]\parskip2pt
\item 
Assume that $f^0_i$ crosses $f^1_j$ for some $j\leq i$, and that $i-j$ is minimized with respect to that.
The minimality assumption implies that $f^1_j$ crosses $P_0$ in the edge of $t^0_{i-1}$.
We~may ``slide'' the vertex $r_0^{2k+i}$ along the drawing of $P_0$ towards $r_0^{2k+i-1}$, such that the crossing of $f^1_j$ with
$P_0$ changes to the edge of weight coefficient $t^0_{i}$.
This increases the crossing~weight on $P_0$ by $(t^0_{i}-t^0_{i-1})\omega^{30+30}\leq\frac{4k}{5k+7}\omega^{60}<(1-\frac1{5})\omega^{60}$
(resp., the same expression with $\omega^{65}$ if $f^1_j=b_3c_3$).
On the other hand, this move saves $\omega^{60}$ (resp., $\omega^{65}$) weight of crossing between $f^0_i$ and $f^1_j$, 
hence decreasing the total number of crossings by at least~$\frac15\omega^{60}$.
\item 
A case that $f^0_i$ crosses $f^1_j$ for some $j>i$ is solved analogously, with ``sliding'' the vertex $r_0^{2k+i}$ along $P_0$ towards $r_0^{2k+i+1}$.

\item \label{it:slide-j-i}
For the rest, we consider that no $f^0_{i'}$ is crossing any $f^1_j$.
Assume that $f^0_i$ crosses $P_1$ in the edge of weight coefficient $t^1_{j}$ for $j<i$, and that $i-j$ is maximized with respect to that.
Further assume that $f^1_j\not=b_3c_3$.
The maximality assumption, together with $f^1_j$ not crossing any $f^0_{i'}$, imply that $f^1_j$ crosses $P_0$ in the edge of $t^0_{i+1}$.
We now simultaneously ``slide'' the end of $f^0_i$ along the drawing of $P_0$ to the right and the end of $f^1_j$ along the drawing 
of $P_1$ to the left, such that $f^0_i$ and $f^1_j$ (informally) exchange positions.

On the path $P_1$, we gain crossings of weight $(t^1_{j+1}-t^1_{j})\omega^{30+30}=\frac{2j+3}{5k+7}\omega^{60}$.
On the path~$P_0$, we save crossings of weight $(t^0_{i+1}-t^0_{i})\omega^{30+30}=\frac{2i+2}{5k+7}\omega^{60}$.
And since $i\geq j+1$, we decrease the total crossings by at least $\frac1{5k+7}\omega^{60}$.

\item \label{it:slide-j-i-65}
Under the same initial assumption as in (\ref{it:slide-j-i}.), we consider the subcase that $f^1_j=b_3c_3$.
Then the same modification of the drawing causes the following.
On the path $P_1$, we gain crossings of weight $\frac25\omega^{35+30}$, neglecting the lower order term.
On the path $P_0$, we save crossings of weight $(t^0_{i+1}-t^0_{i})\omega^{30+35}=\frac{2i+2}{5k+7}\omega^{65}$.
Since $i\geq k+1$ in this case, we decrease the total crossings by at least 
$(\frac{2(k+1)+2}{5k+7}-\frac25)\omega^{65}\geq\frac1{5(5k+7)}\omega^{65}$.

\item 
We analogously handle the cases as (\ref{it:slide-j-i}.) and (\ref{it:slide-j-i-65}.) with~$j>i$.
If $f^1_j\not=b_3c_3$, we descrease the total crossings by at least
$(t^1_{j}-t^1_{j-1})\omega^{30+30}-(t^0_{i+1}-t^0_{i})\omega^{30+30}= (\frac{2j+1}{5k+7}-\frac{2i+2}{5k+7})\omega^{60}\geq\frac1{5k+7}\omega^{60}$.
If $f^1_j=b_3c_3$, we have $i\leq k$ and the saving is at least 
$(\frac25-\frac{2k+2}{5k+7})\omega^{65}\geq\frac4{5(5k+7)}\omega^{65}$.
\end{enumerate}
\vspace*{-\baselineskip}
\end{proof}
\fi

\ifx\proof\inlineproof
\begin{proof}
\else
\begin{proof}[Proof (with further details in the Appendix)]
\fi
As for the first part of the statement, we organize arguments leading to each term of the formula for $\gamma_\omega(k)$ 
stepwise from higher to lower order terms of edge weights.
The proof steps, with details of \eqref{it:P2above} and \eqref{it:rbordering} skipped till later parts of the proof, are as follows:

\begin{enumerate}[(1)]\parskip1ex%
\item\label{it:topbound}
The blue path $P_1$ of weight $\geq\!\omega^{49}$ must not cross the red path $P_0$ of weight $\geq\!\omega^{41}$ since~that 
would result in crossings of weight at least $2\omega^{90}>\gamma_\omega^+(k)$, and likewise with $P_1$ and~$C_0$.
So, by the Jordan curve theorem, the two red edges $r_1r_1'$ and $r_3r_3'$ of weight $\omega^{41}-\omega^{40}$ must cross $P_1$, 
contributing crossing weight at least $2\omega^{90}-2\omega^{89}$.
The $4k$ red edges of weight $\omega^{30}$ from $P_0$ to $C_0$ similarly make $4k\omega^{79}$ crossings with~$P_1$.
Further, there are edge-disjoint paths from $b_2$ to $d_1$ of combined weight $\omega^{48}+2\omega^{38}$;
these are formed by the path $b_2c_2+R$ of weight $\omega^{48}$, by a path from $b_2$ through $c_2$, $c_0$ and $d_0$ of weight~$\omega^{38}$,
by a path from $b_2$ through $c_2$, $c_4$ and $d_4$ of weight $\omega^{38}-\omega^{34}$,
and by a path from $b_2$ through $b_3$, $c_3$ and $c_4$ of weight~$\omega^{34}$.
These contribute weight $\omega^{89}+2\omega^{79}$ of crossings~with~$P_0$.
Summing up, we have so far accounted for at least $2\omega^{90}-\omega^{89}+(4k+2)\omega^{79}>\gamma_\omega(k)-2\omega^{76}+\omega^{75}$ enforced crossings.

\item\label{it:P2above}
The next step is to prove that $P_2$ is drawn disjoint from $P_0$ and drawn above it, and all $c_1b_1$, $c_2b_2$, $c_3b_3$ cross~$P_0$.
If, for an illustration, $P_0$ was crossed by both $b_1c_1$ and $b_3c_3$, but not by $b_2c_2$, the lower bound from \eqref{it:topbound} would rise
(thanks to full weight of $b_1c_1$ and $b_3c_3$) to $2\omega^{90}-\omega^{89}+(4k+2)\omega^{79}+2\omega^{76}\geq\gamma_\omega(k)+\omega^{75}-\omega^{72}>\gamma_\omega^+(k)$, which is impossible.
If neither of $b_1c_1$, $b_3c_3$ crosses $P_0$, then we (briefly) get subpaths of $P_1$ and $P_2$ of combined weight
$2\frac25\omega^{35}+2\frac45\omega^{35}=(2+\frac25)\omega^{35}$ which cross the path $P_0$ or edges $r_1r_1'$, $r_3r_3'$,
and this contributes an additional weight $(2+\frac25)(\omega^{76}-\omega^{75})$ of crossings, again exceeding $\gamma_\omega^+(k)$.
We leave the fine details and subcases for a later part of the proof.

\item\label{it:thirdx}
Using \eqref{it:P2above}, we may count crossings of $P_0$ with the edges $c_1b_1$, $c_2b_2$, $c_3b_3$ and
the $4k$ blue edges of weight $\omega^{30}$ between $P_1$ and $P_2$.
At this point, our lower bound gets to $2\omega^{90}-\omega^{89}+(4k+2)\omega^{79}+2\omega^{76}-\omega^{75}+4k\omega^{71}>\gamma_\omega^+(k)-\omega^{66}$.

\begin{figure}[t]
\hspace*{-1ex}%
\begin{tikzpicture}[scale=4.9]\small
\tikzstyle{every node}=[draw, color=black, thick, shape=circle, inner sep=1.3pt, fill=blue]
\node[label=below:$\!b_2\!$] (bM) at (-0.13,-1.0) {};
\node[label=below:$b_4\!\!$,label=right:$~~P_1$] (bN) at (2.42,-1.0) {};
\node[label=below:$\!\!b_3$] (bR) at (1.2,-1.0) {};
\node[label=left:$c_2$, label=above:\color{blue}\hspace*{-5ex}$\omega^{48}\!+\!2\omega^{38}\!-\!\omega^{34}$\hspace*{-5ex}] (bbM) at (-0.13,-0.1) {};
\node[label=right:$c_3$, label=above:\color{blue}$~\omega^{35}$] (bbR) at (1.2,-0.1) {};
\tikzstyle{every node}=[draw, color=black!50!blue, shape=circle, inner sep=0.85pt, fill=blue]
\node (b1') at (2.2,-1.0) {}; \node (b2') at (1.95,-1.0) {}; \node (b3') at (1.7,-1.0) {}; \node (b4') at (1.45,-1.0) {};
\node (b5') at (0.95,-1.0) {}; \node (b6') at (0.7,-1.0) {}; \node (b7') at (0.45,-1.0) {}; \node (b8') at (0.18,-1.0) {};
\tikzstyle{every path}=[draw, color=blue!80!black, ultra thick]
\draw (bM)--(bN) (bM)--(bbM);
\tikzstyle{every path}=[draw, color=blue, very thick]
\draw (bR)--(bbR);
\tikzstyle{every path}=[draw, color=blue, fill=blue, thick]
\draw (b1')--(2.2,-0.25) node[draw=none,fill=none, label=above:$\omega^{30}$] {};
\draw (b2')--(1.95,-0.25) node[draw=none,fill=none, label=above:$\omega^{30}$] {};
\draw (b3')--(1.7,-0.25) node[draw=none,fill=none, label=above:$\omega^{30}$] {};
\draw (b4')--(1.45,-0.25) node[draw=none,fill=none, label=above:$\omega^{30}$] {};
\draw (b5')--(0.95,-0.25) node[draw=none,fill=none, label=above:$\omega^{30}$] {};
\draw (b6')--(0.7,-0.25) node[draw=none,fill=none, label=above:$\omega^{30}$] {};
\draw (b7')--(0.45,-0.25) node[draw=none,fill=none, label=above:$\omega^{30}$] {};
\draw (b8')--(0.18,-0.25) node[draw=none,fill=none, label=above:$\omega^{30}$] {};

\tikzstyle{every node}=[draw, color=black, thick, shape=circle, inner sep=1.2pt, fill=red!80!black]
\node[label=right:$~P_0$, label=above:$r_4$] (rN) at (2.42,-0.4) {};
\node[label=above:$r_1$] (rL) at (-0.3,-0.4) {}; \node[label=above:$r_3$] (rR) at (0,-0.4) {};
\node[label=left:$r_2$] (rB) at (-0.13,-1.8) {};
\node[label=above:\color{red!80!black}\hspace*{-2ex}$\omega^{41}-\omega^{40}\qquad\qquad$] (rrR) at (0,-1.6) {};
\tikzstyle{every node}=[draw, color=black!50!red, shape=circle, inner sep=0.85pt, fill=red!90!black]
\node[label=above:$r_0^{2k+8}\!\!\!\!\!\!$] (r1') at (2.,-0.4) {}; \node[label=above:$r_0^{2k+7}\!\!\!\!\!\!$] (r2') at (1.75,-0.4) {}; 
\node[label=above:$r_0^{2k+6}\!\!\!\!\!\!$] (r3') at (1.5,-0.4) {}; \node[label=above:$r_0^{2k+5}\!\!\!\!\!\!$] (r4') at (1.25,-0.4) {};
\node[label=above:$r_0^{2k+4}\!\!\!\!\!\!$] (r5') at (1.,-0.4) {}; \node[label=above:$r_0^{2k+3}\!\!\!\!\!\!$] (r6') at (0.75,-0.4) {}; 
\node[label=above:$r_0^{2k+2}\!\!\!\!\!\!$] (r7') at (0.5,-0.4) {}; \node[label=above:$r_0^{2k+1}\!\!\!\!\!$] (r8') at (0.25,-0.4) {};

\tikzstyle{every path}=[draw, color=red!80!black, ultra thick]
\draw (rL)--(rR)--(rN) ;
\draw (rrR) to[bend left] (rB) (rrR)--(rR);
\tikzstyle{every path}=[draw, color=red!90!black, thick]
\tikzstyle{every node}=[draw, color=black!40!red, shape=circle, inner sep=0.6pt, fill=red!80!black]
\draw (r1')-- (2,-1.6) ;
\draw (r2')-- (1.75,-1.6) ;
\draw (r3')-- (1.5,-1.6) ;
\draw (r4')-- (1.25,-1.6) ;
\draw (r5')-- (1,-1.6) node[draw=none,fill=none, label=below:$~\ldots~$] {};
\draw (r6')-- (0.75,-1.6) node[draw=none,fill=none, label=below:$\omega^{30}$] {};
\draw (r7')-- (0.5,-1.6) node[draw=none,fill=none, label=below:$\omega^{30}$] {};
\draw (r8')-- (0.25,-1.6) node[draw=none,fill=none, label=below:$\omega^{30}$] {};

\tikzstyle{every node}=[draw, color=blue!80!black, draw=none,fill=none, rotate=90]
\node at (0.08,-1.13) {$\omega^{49}\to$};
\node at (0.35,-1.25) {$\omega^{49}\!\!+\frac3{27}\omega^{30}\to$};
\node at (0.6,-1.25) {$\omega^{49}\!\!+\frac8{27}\omega^{30}\to$};
\node at (0.85,-1.25) {$\omega^{49}\!\!+\frac{15}{27}\omega^{30}\to$};
\node at (1.08,-1.25) {$\omega^{49}\!\!+\frac{24}{27}\omega^{30}\to$};
\node at (1.35,-1.36) {$\omega^{49}\!\!+\!\frac25\omega^{35}\!\!+\frac{35}{27}\omega^{30}\to$};
\node at (1.6,-1.36) {$\omega^{49}\!\!+\!\frac25\omega^{35}\!\!+\frac{48}{27}\omega^{30}\to$};
\node at (1.85,-1.36) {$\omega^{49}\!\!+\!\frac25\omega^{35}\!\!+\frac{63}{27}\omega^{30}\to$};
\node at (2.1,-1.36) {$\omega^{49}\!\!+\!\frac25\omega^{35}\!\!+\frac{80}{27}\omega^{30}\to$};
\node at (2.3,-1.36) {$\omega^{49}\!\!+\!\frac25\omega^{35}\!\!+\frac{99}{27}\omega^{30}\to$};

\tikzstyle{every node}=[draw, color=red!80!black, draw=none,fill=none, rotate=90]
\node at (-0.23,-0.53) {$\omega^{41}\to$};
\node at (0.1,-0.65) {$\omega^{41}\!\!+\frac{2}{27}\omega^{30}\to$};
\node at (0.35,-0.65) {$\omega^{41}\!\!+\frac{6}{27}\omega^{30}\to$};
\node at (0.6,-0.65) {$\omega^{41}\!\!+\frac{12}{27}\omega^{30}\to$};
\node at (0.85,-0.65) {$\omega^{41}\!\!+\frac{20}{27}\omega^{30}\to$};
\node at (1.1,-0.65) {$\omega^{41}\!\!+\frac{30}{27}\omega^{30}\to$};
\node at (1.35,-0.65) {$\omega^{41}\!\!+\frac{42}{27}\omega^{30}\to$};
\node at (1.6,-0.65) {$\omega^{41}\!\!+\frac{56}{27}\omega^{30}\to$};
\node at (1.85,-0.65) {$\omega^{41}\!\!+\frac{72}{27}\omega^{30}\to$};
\node at (2.1,-0.65) {$\omega^{41}\!\!+\frac{90}{27}\omega^{30}\to$};

\begin{scope}[on background layer]
\draw[black!10!white, solid, line width=2mm] (0,-0.4)--(2.42,-0.4);
\draw[black!10!white, solid, line width=2mm] (0.18,-1)--(2.42,-1);
\end{scope}
\end{tikzpicture}

\caption{A detail of the ``ordering'' of vertices on the paths $P_0$ and $P_1$ of the frame $F_k$ from \Cref{fig:Fk-all}, where~$k=4$.
	Only the half to the right of $b_2c_2$ is shown, and the other half is symmetric.
	The edges of $P_0$, going from $r_1$ on the left to $r_4$ on the right, have weights $\omega^{41}+\frac{i(i+1)}{5k+7}\omega^{30}$ for $i=0,1,\ldots,2k+1$.
	The edges of $P_1$, going from $b_2$ on the left to $b_4$ on the right, have weights $\omega^{49}+\frac{i(i+2)}{5k+7}\omega^{30}$ for $i=0,1,\ldots,k$
		and $\omega^{49}+\frac25\omega^{35}+\frac{i(i+2)}{5k+7}\omega^{30}$ for $i=k+1,\ldots,2k+1$.
	Using straightforward calculus, one can compute that ``sliding'' vertices of $P_0$ across the vertical edge $b_3c_3$ (in any direction) would increase the weight of crossings by $\geq\frac1{5(5k+7)}\omega^{65}$,
	and ``sliding'' with respect to other depicted vertical blue edges would increase the weight by $\geq\frac1{5k+7}\omega^{60}$.
}\label{fig:Fk-order}
\end{figure}

\item\label{it:rbordering}
Next comes the crucial step of the reduction -- at cost of additional $c_1(k)\omega^{65}+c_2(k)\omega^{60}$ weight of crossings,
we ``order'' the vertical blue edges which stretch between the paths $P_1$ and $P_2$ as alternating with the internal vertices of the red path~$P_0$.
This rather long technical step is detailed at the end of the proof, and here we only outline its core idea which is based on simple calculus as follows.
Consider an arbitrary quadratic polynomial $p(x)$ which is increasing on $\mathbb R^+$.
Then, for any $a>0$, the minimum of the function $p(x)+p(y)$ conditioned by $x+y=a$ is attained at $x=y$, that is when~$x=a/2$.
A slight adjustment (suitable for integer values of~$x$) of this idea is used to determine fine weights of the edges of the paths $P_0$ and $P_1$
where, essentially, $x$ means the index of an edge of $P_0$ and $y$ that of an edge of~$P_1$.
Such fine weights then enforce precise mutual positions of the edges of $P_0$ and~$P_1$, and in turn the desired alternating ordering of the vertical red and blue edges incident to $P_0$ and $P_1$.
A closer idea of this principle can be obtained by looking at an example of concrete edge weights in \Cref{fig:Fk-order}.

\item\label{it:redQ} 
Points \eqref{it:topbound}--\eqref{it:rbordering} together imply that the weight of the crossings in $(F_k,B)$ is at least $\gamma_\omega(k)-3k\omega^{52}$.
So, none of the paths of $\ca Q$ may cross $P_1$ since that would add $\omega^{49+4}=\omega^{53}$.
Then each of the paths of $\ca Q$ crosses the blue edge $c_2b_2$ of weight $\omega^{48}+2\omega^{38}-\omega^{34}$,
or the path $R$ plus two sections of the path $P_2$ of combined weight $\omega^{48}+2\omega^{38}$.
Since $2k$ of the paths of $\ca Q$ are of weight $\omega^4$ and $k$ of them of weight $\omega^4-1$, by Theorem~\ref{thm:anchorNPHrefin}(a),
we have at least $(\omega^{48}+2\omega^{38}-\omega^{34})(3k\omega^4-k)\geq 3k\omega^{52}-k\omega^{48}+6k\omega^{42}-5k\omega^{38}$ more crossings from that.
Moreover, each of the paths of $\ca Q$ crosses sections of $P_2$ between $c_0$--$c_1$ and $c_3$--$c_4$ of weight~$\frac45\omega^{35}$
\smallskip
(plus the edges $b_1c_1$ and $b_3c_3$), contributing another $\frac45\omega^{35}(3k\omega^4-k)=\frac{12}5k\omega^{39}-\frac45k\omega^{35}$.
\smallskip
All these together raise the lower bound to at least~$\gamma_\omega(k)-5k\omega^{38}-\frac45k\omega^{35}\geq\gamma_\omega^+(k)-(5k+1)\omega^{38}$.

\item\label{it:gregion}
Assume that some of the internal vertices of a path in $\ca Q$ lie outside of the gadget region (informally, below $P_2$).
That would force an additional crossing between such a path and $c_1b_1$ or $c_3b_3$ of weight $\geq\!(1-\frac45)\omega^{35+4}$
(or with e.g.~$P_2$ of even higher weight), exceeding~$\gamma_\omega^+(k)$.
Therefore, every path $Q\in\ca Q$ crosses the path $P_2$ in the two end-edges of~$Q$ which are of weight $\omega^4$ by the condition in Theorem~\ref{thm:anchorNPHrefin}(a),
and the weights of the edges of $P_2$ crossed by $Q$ are $\omega^{38}$ and $\omega^{38}+\frac45\omega^{35}$, respectively.
Additionally, all paths of $\ca Q$ cross the path $R$ of weight $\omega^{48}$ at least with their minimum edge weights $\omega^4-1$~or~$\omega^4$.
All this improves the estimate on the crossings carried by the paths of $\ca Q$ from \eqref{it:redQ} to
$\geq\omega^{48}(3k\omega^4-k)+(2\omega^{38}+\frac45\omega^{35})\cdot3k\omega^4= 3k\omega^{52}-k\omega^{48}+6k\omega^{42}+\frac{12}5k\omega^{39}$,
which raises our lower bound to the desired value~$\gamma_\omega(k)$.
\end{enumerate}

A drawing with $\gamma_\omega(k)$ crossings is as in \Cref{fig:Fk-all} with details as in \Cref{fig:Fk-order}.
Hence, we have finished a proof of $\cra(F_k,B)= \gamma_\omega(k)$.
Furthermore, if any of the assumptions of the previous analysis was violated, the crossing number would exceed $\gamma_\omega^+(k)$,
and any one additional crossing not sketched in \Cref{fig:Fk-all} and not being only between paths of $\ca Q$
would add weight of at least $\omega^{30+4}-\omega^{30}$, and again $\gamma_\omega(k)+\omega^{34}-\omega^{30}>\gamma_\omega^+(k)$, 
which confirms the rest of Lemma~\ref{lem:framecr}.

\medskip
We continue with additional proof details of the two sketched points.
\begin{claimproof}[Proof of \eqref{it:P2above}]
Once we prove that all three edges $c_1b_1$, $c_2b_2$, $c_3b_3$ cross~$P_0$, we raise the lower bound from \eqref{it:topbound}
to at least $2\omega^{90}-\omega^{89}+(4k+2)\omega^{79}+2\omega^{76}-\omega^{75}>\gamma_\omega^+(k)-\omega^{72}$, and hence $P_2$ could not cross $P_0$ and had to be above it.
It thus suffices to analyze all possibilities that some of $c_1b_1$, $c_2b_2$, $c_3b_3$ do not cross~$P_0$.
\begin{enumerate}[i.]
\item Neither of $c_1b_1$, $c_3b_3$ cross~$P_0$.
Then (regardless of $c_2b_2$) the weight of crossings of order $\omega^{76}$ sums to at least $2\cdot\frac45\omega^{76}$ between $P_0$ and $P_2$, precisely, with the sections of $P_2$ between $c_0$--$c_1$ and $c_3$--$c_4$.
In addition to that, we have a path from $b_0$ to $b_4$, using the edges $b_1c_1$, $b_3c_3$ and the section of $P_2$ between $c_1$--$c_3$, of weight $\frac25\omega^{35}$.
This path must cross twice the subgraph formed by $P_0^+:=P_0\cup\{r_1r_1',r_3r_3'\}$ and, importantly, the possible crossing(s) with $P_0$ is in addition 
to the crossings between $P_0$ and $P_2$ counted in \eqref{it:topbound} if $b_2c_2$ does not cross $P_0$.
So, this adds (neglecting the lower-order terms) at least $2\cdot\frac25\omega^{76}$ more crossings between $P_0^+$ and $P_2$.
Altogether, with the lower bound of \eqref{it:topbound}, we get at least 
$2\omega^{90}-\omega^{89}+(4k+2)\omega^{79}+\frac{12}{5}\omega^{76}-\omega^{75}>\gamma_\omega^+(k)$ crossings, which is impossible.
\item Exactly one of $c_1b_1$, $c_3b_3$ crosses~$P_0$.
We reuse ``one half'' of the argument in (i.) plus the weight of the crossing of one of $c_1b_1$, $c_3b_3$ with $P_0$ to derive an analogous contradiction.
\item Both of $c_1b_1$, $c_3b_3$ cross~$P_0$, but $b_2c_2$ does not.
Then the lower bound from the arguments of \eqref{it:topbound} can be slightly improved, since the path $R$ and twice the path $P_2$ cross $P_0$,
and in addition to that, we count the crossings of $c_1b_1$ and $c_3b_3$ with~$P_0$.
So, we improve it to at least $2\omega^{90}-\omega^{89}+(4k+2)\omega^{79}+2\omega^{76}>\gamma_\omega(k)+\omega^{75}-\omega^{72}>\gamma_\omega^+(k)$, which is again impossible.
\end{enumerate}
\vspace*{-\baselineskip}\end{claimproof}

\ifx\proof\inlineproof
\begin{claimproof}[Proof of \eqref{it:rbordering}]
We first note that the graph $(F_k,B)$ is symmetric along the vertical axis, except the paths of~$\ca Q$.
Since the paths of $\ca Q$ are handled only after point \eqref{it:rbordering}, we can now assume full symmetry and prove the claim only for the right-hand side
of $(F_k,B)$, as depicted in \Cref{fig:Fk-order}, and then multiply the number of crossings by~$2$.

We recapitulate where we stand with our drawing of $(F_k,B)$ after steps \eqref{it:topbound}--\eqref{it:thirdx}.
We have got pairwise noncrossing cycle $C_0$ and paths $P_1$, $P_0$ and $P_2$ drawn in this order bottom up (\Cref{fig:Fk-all}).
All crossings of weight of order $\omega^{66}$ and higher have already been counted to equality with~$\gamma_\omega(k)$.
There are two edges of weight $\omega^{41}-\omega^{40}$ between $C_0$ and $P_0$ crossing $P_1$, and $4k$ such edges of weight $\omega^{30}$.
There is an edge of weight $\omega^{48}+2\omega^{38}-\omega^{34}$ and two edges of weight $\omega^{35}$ between $P_1$ and $P_2$ crossing $P_0$,
and again $4k$ such edges of weight $\omega^{30}$.
Besides the already counted weights, these listed edges contribute crossings of weights of order only $\omega^{65}$ and $\omega^{60}$
(including their possible mutual crossings).

Our proof strategy is the following; we will show a concrete drawing, called the \emph{normal drawing}, in which the above described crossings contribute exactly as expected 
in the formula for~$\gamma_\omega(k)$ in the part $c_1(k)\omega^{65}+c_2(k)\omega^{60}$ (the drawing in \Cref{fig:Fk-order}),
and then we will argue that in any drawing in which the above listed edges (the vertical blue and red ones of weights $\omega^{35}$and $\omega^{30}$)
are not as in our normal drawing, we can decrease the total crossing number by $\Omega(\omega^{60})$.
This possible drop in the crossing number in turn certifies that the arbitrary considered drawing would exceed $\gamma_\omega^+(k)$ crossings,
and hence is impossible in our claim.
In this setup of a proof, it also comes for free that all crossings potentially occuring in $(F_k,B)$, 
which are not counted prior to this point and are not among the crossings listed above, are of weight of order strictly less than~$\omega^{60}$.

The weights of the edges of $P_0$ and $P_1$ are precisely as follows (\Cref{fig:Fk-order}):
\begin{itemize}\parskip2pt
\item For $P_0$, $V(P_0)=(r_0,r_0^1,\ldots,r_0^{2k},r_1,r_3,r_0^{2k+1},\ldots,r_0^{4k},r_4)$,
\smallskip
the weight of $r_1r_3$ is exactly $\omega^{41}$, the weight of $r_0^{2k}r_1$ and of $r_3r_0^{2k+1}$ is $\omega^{41}\!+\frac{2}{5k+7}\omega^{30}$,
the weight of $r_0^{2k+i-1}r_0^{2k+i}$ and of $r_0^{2k-i+1}r_0^{2k-i}$ is $\omega^{41}\!+\frac{i(i+1)}{5k+7}\omega^{30}$ for $i=2,\ldots,2k$,
and the weight of $r_0r_0^1$ and of $r_0^{4k}r_4$ is $\omega^{41}\!+\frac{(2k+1)(2k+2)}{5k+7}\omega^{30}$.

\item For $P_1$, $V(P_1)=(b_0,b_0^1,\ldots,b_0^{k},b_1,b_0^{k+1},\ldots,b_0^{2k},b_2,b_0^{2k+1},\ldots,b_0^{3k},b_3,b_0^{3k+1},$ $\ldots,b_0^{4k},b_4)$,
we resort to describing the weights only from $b_2$ till $b_4$.
The weight of $b_2b_0^{2k+1}$ is exactly $\omega^{49}$, 
the weight of $b_0^{2k+i}b_0^{2k+i+1}$ is $\omega^{49}\!+\frac{i(i+2)}{5k+7}\omega^{30}$ for $i=1,\ldots,k-1$,
the weight of $b_0^{3k}b_3$ is $\omega^{49}\!+\frac{k(k+2)}{5k+7}\omega^{30}$, 
the weight of $b_3b_0^{3k+1}$ is $\omega^{49}\!+\frac25\omega^{35}\!+\frac{(k+1)(k+3)}{5k+7}\omega^{30}$,
the weight of $b_0^{3k+i}b_0^{3k+i+1}$ is $\omega^{49}\!+\frac25\omega^{35}\!+\frac{(i+k+1)(i+k+3)}{5k+7}\omega^{30}$ for $i=1,\ldots,k-1$,
and the weight of $b_0^{4k}b_4$ is $\omega^{49}\!+\frac25\omega^{35}\!+\frac{(2k+1)(2k+3)}{5k+7}\omega^{30}$.
\end{itemize}

We count the crossings in a normal drawing as in \Cref{fig:Fk-order}:
\begin{itemize}
\item Concerning total crossings of weight of order $\omega^{65}$, 
we have $2k$ red edges of weight $\omega^{30}$ crossing the sections of $P_1$ between $b_0$--$b_1$ and $b_3$--$b_4$,
and two blue edges of weight $\omega^{35}$ crossing $P_0$ in edges of weight $\frac{(k+1)(k+2)}{5k+7}$. In total 
\begin{eqnarray*}
2k\omega^{30}\cdot\frac25\omega^{35}+2\omega^{35}\cdot\frac{(k+1)(k+2)}{5k+7}\omega^{30}=\frac2{5(5k+7)}(30k^2+58k+20)\omega^{65}
 =c_1(k)\omega^{65}.
\end{eqnarray*}
\item Concerning total crossings of weight of order $\omega^{60}$, and considering only the right-hand side as in \Cref{fig:Fk-order},
we get crossings of $2k$ red edges of weight $\omega^{30}$ with the edges of $P_1$ from $b_0^{2k+1}$ to $b_0^{4k}$,
and crossings of $2k$ blue edges of weight $\omega^{30}$ with the edges of $P_0$ from $r_3$ to $r_4$ except with the edge of weight $\frac{(k+1)(k+2)}{5k+7}$
(counted in the previous point). These sum to
\begin{eqnarray*}
\omega^{30}\cdot&&\sum_{i=1}^{2k}\frac{i(i+2)}{5k+7}\omega^{30} + \omega^{30}\cdot\left[\,\sum_{i=1}^{2k+1}\frac{i(i+1)}{5k+7}-\frac{(k+1)(k+2)}{5k+7}\right]\omega^{30}
\\&&=\frac1{3(5k+7)}(16k^3+39k^2+20k)
,\end{eqnarray*}
which multiplied by $2$ gives~$c_2(k)$.
\end{itemize}

Now, we handle an arbitrary drawing of $(F_k,B)$ which conforms to the bounds shown in \eqref{it:topbound}--\eqref{it:thirdx}, 
but is not our normal drawing described by \Cref{fig:Fk-order}.
As argued above, it is enough for this purpose to consider only edges depicted in \Cref{fig:Fk-order}, and know that the depicted vertical
edges indeed cross the horizontal paths $P_1$ and $P_2$ somewhere. 
Crossings of weight of order strictly higher than $\omega^{65}$ are not relevant in this analysis (as they are enforced and have been counted,
and so new such ones cannot even arise), and crossings of weight of order strictly less than $\omega^{60}$ can be ignored at this stage (by the choice of sufficiently large $\omega$).

We denote by $t^0_i=\frac{i(i+1)}{5k+7}$ and $t^1_i=\frac{i(i+2)}{5k+7}$, and refer to \Cref{fig:Fk-order}.
The $i$-th edge of the red path $P_0$, counted from $r_3$ towards $r_4$, is of weight $\omega^{41}\!+t^0_i\omega^{30}$,
and the $(i+1)$-th edge of the blue path $P_1$, counted from $b_2$ towards $b_4$, is of weight $\omega^{49}\!+t^1_i\omega^{30}$,
resp.~of weight $\omega^{49}\!+\frac25\omega^{35}\!+t^1_i\omega^{30}$ if~$i>k$.
We also denote the vertical red edges (of weight $\omega^{30}$) incident to the path $P_0$ by $f^0_1,\ldots,f^0_{2k}$ such that
$f^0_i$ is incident to the vertex $r_0^{2k+i}$, and the vertical blue edges (of weight again~$\omega^{30}$) incident to the path $P_1$ 
by $f^1_1,\ldots,f^1_{2k+1}$ such that $f^1_i$ is incident to the vertex $b_0^{2k+i}$ for $i\leq k$,~ 
$f^1_{k+1}$ is incident to $b_3$, and $f^1_i$ is incident to $b_0^{2k+i-1}$ for $i\geq k+2$.

One can easily check that in our normal drawing, we have, naturally, $f^0_i$ crossing $P_1$ in the edge of weight coefficient $t^1_i$,
~$f^1_j$ crossing $P_0$ in the edge of weight coefficient $t^0_j$, and no $f^0_i$ is crossing any $f^1_j$.
Assume that the drawing of $f^0_i$ violates some of the previous conditions.
\begin{enumerate}[i.]\parskip2pt
\item 
Assume that $f^0_i$ crosses $f^1_j$ for some $j\leq i$, and that $i-j$ is minimized with respect to that.
The minimality assumption implies that $f^1_j$ crosses $P_0$ in the edge of $t^0_{i-1}$.
We~may ``slide'' the vertex $r_0^{2k+i}$ along the drawing of $P_0$ towards $r_0^{2k+i-1}$, such that the crossing of $f^1_j$ with
$P_0$ changes to the edge of weight coefficient $t^0_{i}$.
This increases the crossing~weight on $P_0$ by $(t^0_{i}-t^0_{i-1})\omega^{30+30}\leq\frac{4k}{5k+7}\omega^{60}<(1-\frac1{5})\omega^{60}$
(resp., the same expression with $\omega^{65}$ if $f^1_j=b_3c_3$).
On the other hand, this move saves $\omega^{60}$ (resp., $\omega^{65}$) weight of crossing between $f^0_i$ and $f^1_j$, 
hence decreasing the total number of crossings by at least~$\frac15\omega^{60}$.
\item 
A case that $f^0_i$ crosses $f^1_j$ for some $j>i$ is solved analogously, with ``sliding'' the vertex $r_0^{2k+i}$ along $P_0$ towards $r_0^{2k+i+1}$.

\item \label{it:slide-j-i}
For the rest, we consider that no $f^0_{i'}$ is crossing any $f^1_j$.
Assume that $f^0_i$ crosses $P_1$ in the edge of weight coefficient $t^1_{j}$ for $j<i$, and that $i-j$ is maximized with respect to that.
Further assume that $f^1_j\not=b_3c_3$.
The maximality assumption, together with $f^1_j$ not crossing any $f^0_{i'}$, imply that $f^1_j$ crosses $P_0$ in the edge of $t^0_{i+1}$.
We now simultaneously ``slide'' the end of $f^0_i$ along the drawing of $P_0$ to the right and the end of $f^1_j$ along the drawing 
of $P_1$ to the left, such that $f^0_i$ and $f^1_j$ (informally) exchange positions.

On the path $P_1$, we gain crossings of weight $(t^1_{j+1}-t^1_{j})\omega^{30+30}=\frac{2j+3}{5k+7}\omega^{60}$.
On the path~$P_0$, we save crossings of weight $(t^0_{i+1}-t^0_{i})\omega^{30+30}=\frac{2i+2}{5k+7}\omega^{60}$.
And since $i\geq j+1$, we decrease the total crossings by at least $\frac1{5k+7}\omega^{60}$.

\item \label{it:slide-j-i-65}
Under the same initial assumption as in (\ref{it:slide-j-i}.), we consider the subcase that $f^1_j=b_3c_3$.
Then the same modification of the drawing causes the following.
On the path $P_1$, we gain crossings of weight $\frac25\omega^{35+30}$, neglecting the lower order term.
On the path $P_0$, we save crossings of weight $(t^0_{i+1}-t^0_{i})\omega^{30+35}=\frac{2i+2}{5k+7}\omega^{65}$.
Since $i\geq k+1$ in this case, we decrease the total crossings by at least 
$(\frac{2(k+1)+2}{5k+7}-\frac25)\omega^{65}\geq\frac1{5(5k+7)}\omega^{65}$.

\item 
We analogously handle the cases as (\ref{it:slide-j-i}.) and (\ref{it:slide-j-i-65}.) with~$j>i$.
If $f^1_j\not=b_3c_3$, we descrease the total crossings by at least
$(t^1_{j}-t^1_{j-1})\omega^{30+30}-(t^0_{i+1}-t^0_{i})\omega^{30+30}= (\frac{2j+1}{5k+7}-\frac{2i+2}{5k+7})\omega^{60}\geq\frac1{5k+7}\omega^{60}$.
If $f^1_j=b_3c_3$, we have $i\leq k$ and the saving is at least 
$(\frac25-\frac{2k+2}{5k+7})\omega^{65}\geq\frac4{5(5k+7)}\omega^{65}$.
\end{enumerate}
\vspace*{-\baselineskip}\end{claimproof}
\else 

\begin{claimproof}[Proof of \eqref{it:rbordering}\,*]
Due to space restrictions, we can include only a sketch of the full argument, which can thus be found in the Appendix.

We recapitulate where we stand with our drawing of $(F_k,B)$ after steps \eqref{it:topbound}--\eqref{it:thirdx}.
We have got pairwise noncrossing cycle $C_0$ and paths $P_1$, $P_0$ and $P_2$ drawn in this order bottom up (\Cref{fig:Fk-all}).
All crossings of weight of order $\omega^{66}$ and higher have already been counted to equality with~$\gamma_\omega(k)$.
There are two edges of weight $\omega^{41}-\omega^{40}$ between $C_0$ and $P_0$ crossing $P_1$, and $4k$ such edges of weight $\omega^{30}$.
There is an edge of weight $\omega^{48}+2\omega^{38}-\omega^{34}$ and two edges of weight $\omega^{35}$ between $P_1$ and $P_2$ crossing $P_0$,
and again $4k$ such edges of weight $\omega^{30}$.

To continue, the weights of the edges of $P_0$ and $P_1$ are precisely as follows (\Cref{fig:Fk-order}):
\begin{itemize}\parskip2pt
\item For $P_0$, $V(P_0)=(r_0,r_0^1,\ldots,r_0^{2k},r_1,r_3,r_0^{2k+1},\ldots,r_0^{4k},r_4)$,
\smallskip
the weight of $r_1r_3$ is exactly $\omega^{41}$, the weight of $r_0^{2k}r_1$ and of $r_3r_0^{2k+1}$ is $\omega^{41}\!+\frac{2}{5k+7}\omega^{30}$,
the weight of $r_0^{2k+i-1}r_0^{2k+i}$ and of $r_0^{2k-i+1}r_0^{2k-i}$ is $\omega^{41}\!+\frac{i(i+1)}{5k+7}\omega^{30}$ for $i=2,\ldots,2k$,
and the weight of $r_0r_0^1$ and of $r_0^{4k}r_4$ is $\omega^{41}\!+\frac{(2k+1)(2k+2)}{5k+7}\omega^{30}$.

\item For $P_1$, $V(P_1)=(b_0,b_0^1,\ldots,b_0^{k},b_1,b_0^{k+1},\ldots,b_0^{2k},b_2,b_0^{2k+1},\ldots,b_0^{3k},b_3,b_0^{3k+1},$ $\ldots,b_0^{4k},b_4)$,
we resort to describing the weights only from $b_2$ till $b_4$.
The weight of $b_2b_0^{2k+1}$ is exactly $\omega^{49}$, 
the weight of $b_0^{2k+i}b_0^{2k+i+1}$ is $\omega^{49}\!+\frac{i(i+2)}{5k+7}\omega^{30}$ for $i=1,\ldots,k-1$,
the weight of $b_0^{3k}b_3$ is $\omega^{49}\!+\frac{k(k+2)}{5k+7}\omega^{30}$, 
the weight of $b_3b_0^{3k+1}$ is $\omega^{49}\!+\frac25\omega^{35}\!+\frac{(k+1)(k+3)}{5k+7}\omega^{30}$,
the weight of $b_0^{3k+i}b_0^{3k+i+1}$ is $\omega^{49}\!+\frac25\omega^{35}\!+\frac{(i+k+1)(i+k+3)}{5k+7}\omega^{30}$ for $i=1,\ldots,k-1$,
and the weight of $b_0^{4k}b_4$ is $\omega^{49}\!+\frac25\omega^{35}\!+\frac{(2k+1)(2k+3)}{5k+7}\omega^{30}$.
\end{itemize}

Our proof strategy is the following; we give a concrete drawing, called the \emph{normal drawing}, in which the above described crossings contribute exactly as expected 
in the formula for~$\gamma_\omega(k)$ in the part $c_1(k)\omega^{65}+c_2(k)\omega^{60}$ (the drawing in \Cref{fig:Fk-order}),
and then we argue that in any drawing in which the above listed edges (the vertical blue and red ones of weights $\omega^{35}$and $\omega^{30}$)
are not as in our normal drawing, we can decrease the total crossing number by $\Omega(\omega^{60})$.
This possible drop in the crossing number in turn certifies that the arbitrary considered drawing would exceed $\gamma_\omega^+(k)$ crossings,
and hence is impossible in our claim.

In the first part of the outline, we routinely sum the crossings in the normal drawing as
\begin{eqnarray*}~\quad
2k\omega^{30}\cdot\frac25\omega^{35}+2\omega^{35}\cdot\frac{(k+1)(k+2)}{5k+7}\omega^{30}=\frac2{5(5k+7)}(30k^2+58k+20)\omega^{65}
 =c_1(k)\omega^{65}
\end{eqnarray*}
and\,/\,plus
\begin{eqnarray*}
2\omega^{30}\cdot\!\!\!\!\!&&\sum_{i=1}^{2k}\frac{i(i+2)}{5k+7}\omega^{30} + 2\omega^{30}\cdot\left[\,\sum_{i=1}^{2k+1}\frac{i(i+1)}{5k+7}-\frac{(k+1)(k+2)}{5k+7}\right]\omega^{30}
\\&&=\frac2{3(5k+7)}(16k^3+39k^2+20k) =c_2(k)\omega^{60}
.\end{eqnarray*}

For the second part of the outline, 
we handle an arbitrary drawing of $(F_k,B)$ which conforms to the bounds shown in \eqref{it:topbound}--\eqref{it:thirdx}, 
but is not our normal drawing described by \Cref{fig:Fk-order}.
As argued above, it is enough for this purpose to consider only edges depicted in \Cref{fig:Fk-order}, and know that the depicted vertical
edges indeed cross the horizontal paths $P_1$ and $P_2$ somewhere. 
Crossings of weight of order less than $\omega^{60}$ can be ignored at this stage.

The informal conclusion now is, that if the vertical red and blue edges in \Cref{fig:Fk-order} do not cross the paths
$P_1$ and $P_0$ in the exact order as depicted, then one can appropriately ``slide'' one of the wrongly ordered
vertical edges to the left or right, resulting in the claimed decrease of the number of crossings by order $\Omega(\omega^{60})$.
This is a routine, although lengthy, calculus exercise detailed in the Appendix.
\end{claimproof}
\fi
The proof of Lemma~\ref{lem:framecr} is finished.
\end{proof}

We are now ready to finish the proof of our main result.

\begin{proof}[Proof of \Cref{thm:main3}]
According to the (hard) instance $(H,A)$ in Theorem~\ref{thm:anchorNPHrefin}, we choose $k$ such that $|A|=4k$, and for
a ``sufficiently large'' integer $\omega$ (see below), we set the weight $w=\omega^4$ in the statement of Theorem~\ref{thm:anchorNPHrefin}.
We make the union $(\bar H,B):=H\cup(F_k,B)$ such that the path $R$ and the paths in $\ca Q$ get identified between $H$ and $F_k$.
Precisely, the red anchors in $A_1\subseteq A$ are identified in the natural order with the internal vertices 
of the path $P_0\subseteq F_k$ except $r_1,r_3$.
The blue anchors in $A_2\subseteq A$ are identified in the natural order with the internal vertices 
of the path $P_2\subseteq F_k$, except the two neighbours of $c_2$, \mbox{and with the vertex~$d_2$.}

The core parameter $\omega$ of the reduction is handled precisely as follows;
denoting by $m$ the number of edges of the simplification of $\bar H$ (i.e., counting parallel edges as one), we choose $\omega>m^2$
and such that the defined weights are all integers, for instance, that $\omega$ is a multiple of $5(5k+7)$.
This choice means that $\omega$ is larger than the largest possible number of edge crossings (but not the summed weight of them)
in an optimal drawing of $(\bar H,B)$, and so for every $t$, a crossing of weight $\omega^{t+1}$ is more than any sum
of crossings of weights at most $\omega^t$ in the expected solution, as needed in our reduction.

If $\cra(H,A)\leq r$, then the witness drawing of $(H,A)$ can be trivially combined with that of $(F_k,B)$ in \Cref{fig:Fk-all},
giving $\cra(\bar H,B)\leq \gamma_\omega(k)+r-(3k\omega^{52}-k\omega^{48})$ by Lemma~\ref{lem:framecr} 
(we subtract the crossings between $R$ and paths of $\ca Q$ which are counted twice).

On the other hand, if $\cra(\bar H,B)\leq \gamma_\omega(k)+s$, then $s=\ca O(\omega^{16+16})<\omega^{33}$ since all edges of $H-E(R)$ are of weight
$\leq w^4=\omega^{16}$, and so we have a drawing of $(\bar H,B)$ whose restriction to $F_k$ conforms to Lemma~\ref{lem:framecr}.
Since crossings of $\ca Q$ with $R$ contribute $\geq3k\omega^{52}-k\omega^{48}$ by the condition on $\ca Q$ in Lemma~\ref{lem:framecr}, we get that
$\cra(H,A)\leq \cra(\bar H,B)-\gamma_\omega(k)+(3k\omega^{52}-k\omega^{48})\leq s+(3k\omega^{52}-k\omega^{48})$
(in the formula, we add back the crossings between $R$ and paths of $\ca Q$ which still exist in the instance~$(H,A)$).

Therefore, with $r-(3k\omega^{52}-k\omega^{48})=s$, there is a drawing of $(H,A)$ with $r$ crossings if and only if
there is a drawing of $(\bar H,B)$ with $\gamma_\omega(k)+r-(3k\omega^{52}-k\omega^{48})$ crossings.
\end{proof}

\section{Conclusions}\label{sec:conclu}

We have completely answered a question of the computational complexity of the anchored crossing number problem in its perhaps most ``innocent''
looking form, in which the input consists of a disjoint union of anchored planar graphs (the PP anchored crossing number problem).
We have proved that the computational complexity jumps straight from near triviality with two anchors to \NP-hardness with three anchors.

We may also slightly relax the conditions in the PP anchored crossing number instances~$(H,A)$;
instead of requiring $H$ to be a disjoint union of two anchored planar graphs, we only require $H$ to be a union of two
anchored planar graphs which are disjoint except possibly at the anchors.
Then it makes sense to consider less than $3+3=6$ anchors in total, and indeed, this problem stays hard with $5$ anchors
in total, as can be seen from our reduction in \Cref{fig:Fk-all} in which we identify $r_0=b_0$.
We believe one can go down to $4$ or even $3$ anchors in total, but a different reduction would probably be necessary.

\smallskip
Our result closely relates to the analogous question of the crossing number of almost planar graphs.
There we see a slight complexity gap, while for almost planar graphs $G+e$ with $\Delta(G)\leq3$ we know a linear-time algorithm,
our new result implies (Corollary~\ref{cor:main3}) that the latter problem becomes \NP-hard when $G$ has \emph{three} vertices of degree greater than~$3$.
A big question remains about graphs $G$ with \emph{one} or \emph{two} vertices of degree greater than~$3$.
This particular question turns out to be surprisingly difficult (we have tried hard to provide at least a partial answer), 
and we can so far only make a conjecture:
\begin{conjecture}\label{conj:almost12}
Let $G$ be a planar graph such that at most \emph{two vertices} of $G$ are of degree greater than~$3$, and $u,v\in V(G)$.
Then one can compute the crossing number of the almost planar graph~$G+uv$ in polynomial time.
\end{conjecture}

\begin{figure}[t]
$$
\begin{tikzpicture}[scale=1.4]\small
\tikzstyle{every path}=[draw, color=black, thick]
\tikzstyle{every node}=[draw, color=black, fill=black, shape=circle, inner sep=1.1pt]
\draw[fill=white!79!black] (0,0) ellipse (50pt and 40pt);
\draw[fill=white] (0.8,0) to[bend left=20] (1.1,-0.2) to[bend left=99] (-1,-0.6)
	to[bend left=70] (-0.2,-0.3) to[bend right] (0.8,0)
	to[bend right=20] (0.2,0.3) -- (-0.2,0.3) to[bend left=60]
	(-1,0.6) to[bend left=99] (1.1,0.2) to[bend left=20] (0.8,0);
\draw[fill=white] (-0.2,-0.3) -- (0.2,0.3) -- (-0.2,0.3) -- (-0.2,-0.3);
\draw (0.8,0) node (a) {} to[bend left=12] (0.8,-0.5) node[label=below:$\!u$] (b) {} -- (0.5,-0.5) node (c) {} to[bend right=12] (0.8,0);
\draw[color=green!66!black] (c) to[bend left] (-0.2,-0.3) node (d) {} (0.2,0.3) node {} -- (-0.2,0.3) node {};
\draw[color=red!66!black] (b) to[bend right=33] (0.9,0.7) node[label=above:$v\!$] {};
\tikzstyle{every node}=[draw, color=black, fill=none, shape=circle, inner sep=2.4pt]
\node at (-0.2,-0.3) {};  \node at (0.8,0) {};
\end{tikzpicture}
$$
\caption{An example construction of an almost planar graph $G+uv$ (where the edge $uv$ is in red), such that only
	the {two} encircled vertices are of degree higher than~$3$,
	and that the gray regions stand for very dense rigid patches. While $\crg(G+uv)=1$ (just cross the two green edges), 
	the best way of inserting the (red) edge $uv$ into a planar drawing of $G$ can be arbitrarily costly.
}\label{fig:insgapx}
\end{figure}

To slightly demonstrate nontriviality of the problem in Conjecture~\ref{conj:almost12}, we remark that if only one vertex of $G$ is of degree
more than~$3$, the gap between the crossing number and the best insertion of $uv$ into a planar drawing of~$G$
(recall that this gap is null when $\Delta(G)\leq3$ \cite{DBLP:journals/algorithmica/CabelloM11}) can be arbitrarily large for sufficiently high values of $\crg(G+uv)$.
With two vertices of degree more than~$3$, the gap is not even proportional:
\begin{claim}
For any $m$ there is a planar graph $G$ with all vertices except two of degree at most $3$, and $u,v\in V(G)$ (\Cref{fig:insgapx}),
such that $\crg(G+uv)=1$ and there is no planar drawing of $G$ into which the edge $uv$ could be inserted with less than $m$ crossings.
\end{claim}

\smallskip
At last, we can change our viewpoint on the studied problem.
While the results show that the crossing number problem of almost planar graphs is para-\NP-hard when taking
the number of vertices of degree greater than $3$, what about considering the maximum degree as the parameter instead?
\begin{problem}\label{prob:amaxdeg}
Let $G$ be a planar graph such that the maximum degree of $G$ is~$d$, and $u,v\in V(G)$.
What is the parameterized complexity of computing $\crg(G+uv)$ with respect to the parameter~$d$?
\end{problem}

We believe this problem belongs to the class \XP; one possible approach could be to ``guess'' the rotation system of edges
of $G$ in \XP-time, and then, say, reduce the problem to cubic graphs -- unfortunately, this does not preserve planarity
of the modified graph~$G$. We are not aware of any results in the direction of Problem~\ref{prob:amaxdeg},
besides approximations in \cite{DBLP:conf/gd/HlinenyS06,DBLP:journals/algorithmica/CabelloM11}.
However, recall that it is \NP-hard to compute the crossing number of general cubic graphs~\cite{Hlineny06}.

\bibliography{anchor-ap.bib}

\begin{thebibliography}{10}

\bibitem{Cabello13}
Sergio Cabello.
\newblock Hardness of approximation for crossing number.
\newblock {\em Discrete Comput. Geom.}, 49(2):348--358, March 2013.

\bibitem{DBLP:journals/algorithmica/CabelloM11}
Sergio Cabello and Bojan Mohar.
\newblock Crossing number and weighted crossing number of near-planar graphs.
\newblock {\em Algorithmica}, 60(3):484--504, 2011.

\bibitem{DBLP:journals/siamcomp/CabelloM13}
Sergio Cabello and Bojan Mohar.
\newblock Adding one edge to planar graphs makes crossing number and
  1-planarity hard.
\newblock {\em {SIAM} J. Comput.}, 42(5):1803--1829, 2013.

\bibitem{GareyJ83}
Michael~R. Garey and David~S. Johnson.
\newblock Crossing number is {NP-complete}.
\newblock {\em SIAM J. Algebr. Discrete Methods}, 4(3):312--316, September
  1983.

\bibitem{DBLP:journals/algorithmica/GutwengerMW05}
Carsten Gutwenger, Petra Mutzel, and Ren{\'{e}} Weiskircher.
\newblock Inserting an edge into a planar graph.
\newblock {\em Algorithmica}, 41(4):289--308, 2005.

\bibitem{Hlineny06}
Petr Hlin\v{e}n{\'{y}}.
\newblock Crossing number is hard for cubic graphs.
\newblock {\em Journal of Comb. Theory, Ser. B}, 96(4):455--471, 2006.
\newblock \href {https://doi.org/10.1016/j.jctb.2005.09.009}
  {\path{doi:10.1016/j.jctb.2005.09.009}}.

\bibitem{DBLP:conf/compgeom/HlinenyD16}
Petr Hlin\v{e}n{\'{y}} and Marek Dern{\'{a}}r.
\newblock Crossing number is hard for kernelization.
\newblock In {\em SoCG}, volume~51 of {\em LIPIcs}, pages 42:1--42:10. Schloss
  Dagstuhl - Leibniz-Zentrum f{\"{u}}r Informatik, 2016.

\bibitem{DBLP:conf/gd/HlinenyS06}
Petr Hlin\v{e}n{\'{y}} and Gelasio Salazar.
\newblock On the crossing number of almost planar graphs.
\newblock In {\em {GD}}, volume 4372 of {\em Lecture Notes in Computer
  Science}, pages 162--173. Springer, 2006.

\bibitem{DBLP:conf/isaac/HlinenyS15}
Petr Hlin\v{e}n{\'{y}} and Gelasio Salazar.
\newblock On hardness of the joint crossing number.
\newblock In {\em {ISAAC}}, volume 9472 of {\em Lecture Notes in Computer
  Science}, pages 603--613. Springer, 2015.

\bibitem{HopcroftT74}
John Hopcroft and Robert Tarjan.
\newblock Efficient planarity testing.
\newblock {\em J. ACM}, 21(4):549–568, oct 1974.
\newblock \href {https://doi.org/10.1145/321850.321852}
  {\path{doi:10.1145/321850.321852}}.

\bibitem{DBLP:journals/informaticaSI/Mohar06}
Bojan Mohar.
\newblock On the crossing number of almost planar graphs.
\newblock {\em Informatica (Slovenia)}, 30(3):301--303, 2006.

\bibitem{DBLP:journals/algorithmica/PelsmajerSS11}
Michael~J. Pelsmajer, Marcus Schaefer, and Daniel Stefankovic.
\newblock Crossing numbers of graphs with rotation systems.
\newblock {\em Algorithmica}, 60(3):679--702, 2011.

\bibitem{zbMATH00881174}
Adrian Riskin.
\newblock The crossing number of a cubic plane polyhedral map plus an edge.
\newblock {\em Stud. Sci. Math. Hung.}, 31(4):405--413, 1996.

\bibitem{WeiKuanW99}
Shih Wei-Kuan and Hsu Wen-Lian.
\newblock A new planarity test.
\newblock {\em Theoretical Computer Science}, 223(1):179--191, 1999.
\newblock \href {https://doi.org/10.1016/S0304-3975(98)00120-0}
  {\path{doi:10.1016/S0304-3975(98)00120-0}}.

\end{thebibliography}

\end{document}